\documentclass[a4paper,14pt]{article}
\usepackage[english]{babel}
\pdfoutput=1 

\usepackage{jheppub} 

\usepackage[T1]{fontenc} 

\usepackage{bbold}

\usepackage[T1,T2A]{fontenc}
\usepackage[utf8x]{inputenc}

\usepackage{amsmath}
\makeatletter
\newcommand{\doublewidetilde}[1]{{%
  \mathpalette\double@widetilde{#1}%
}}
\newcommand{\double@widetilde}[2]{%
  \sbox\z@{$\m@th#1\widetilde{#2}$}%
  \ht\z@=.9\ht\z@
  \widetilde{\box\z@}%
}
\makeatother

\usepackage{mathrsfs}

\usepackage{xcolor}

\usepackage{comment}

\usepackage{cancel}

\usepackage{cases}

\usepackage{mathtools}

\newcommand{\bi}{\begin{itemize}}
    \newcommand{\ei}{\end{itemize}}
\newcommand{\bea}{\begin{eqnarray}}
    \newcommand{\eea}{\end{eqnarray}}
\newcommand{\bt}{\begin{tabular}}
    \newcommand{\et}{\end{tabular}}
\newcommand{\bc}{\begin{center}}
    \newcommand{\ec}{\end{center}}

\newcommand{\be}{\begin{equation}}
    \newcommand{\ee}{\end{equation}}
\newcommand{\ba}{\begin{array}}
    \newcommand{\ea}{\end{array}}

\def\bbox{{\,\lower0.9pt\vbox{\hrule \hbox{\vrule height 0.2 cm
                \hskip 0.2 cm \vrule height 0.2 cm}\hrule}\,}}
\newcommand{\dsl}{\pa \kern-0.5em /}

\makeatletter \@addtoreset{equation}{section} \makeatother

\def\slashchar#1{\setbox0=\hbox{$#1$}           
    \dimen0=\wd0                                 
    \setbox1=\hbox{/} \dimen1=\wd1               
    \ifdim\dimen0>\dimen1                        
    \rlap{\hbox to \dimen0{\hfil/\hfil}}      
    #1                                        
    \else                                        
    \rlap{\hbox to \dimen1{\hfil$#1$\hfil}}   
    /                                         
    \fi}

\pdfoutput=1

\title{\boldmath  Wess-Zumino gauge for analytic prepotentials of off-shell $\mathcal{N}=2$ supergravity: bosonic sector}

\author[a,b]{Nikita~Zaigraev,}
\author[b,c]{Julia~Zernina}

\affiliation[a]{Bogoliubov Laboratory of Theoretical Physics, JINR,\\141980 Dubna, Moscow region, Russia}
\affiliation[b]{Moscow Institute of Physics and Technology,\\ 141700 Dolgoprudny, Moscow region, Russia}
\affiliation[c]{Institute of Theoretical and Mathematical Physics, Lomonosov Moscow State University,\\ 119991 Moscow, Russia}

\emailAdd{nikita.zaigraev@phystech.edu}
\emailAdd{zernina.iua@phystech.edu}

\abstract{
We present a comprehensive study of the Wess–Zumino gauge for the analytic prepotentials of the off-shell $\mathcal{N}=2$ Weyl and $\mathcal{N}=2$ Einstein supergravities (the version with a non-linear tensor compensator). The field redefinitions, the full transformation laws, and the gauge superparameters that preserve the Wess–Zumino gauge are derived. As a concrete application of the obtained results, we examine the component structure of the free hypermultiplet action coupled to $\mathcal{N}=2$ supergravity. In this work, we restrict our analysis strictly to the bosonic sector.
}

\makeatletter
\gdef\@fpheader{}
\makeatother

\begin{document}

\maketitle
\flushbottom


\section{Introduction}

One of the fundamental problems in supergravity is to find a complete set of unconstrained superfield prepotentials – i.e., the superfields that contain all physical degrees of freedom with the correct gauge transformations without imposing differential constraints by hand \cite{Gates:1983nr, BK, Wess:1992cp}. The difficulty arises as the standard geometric formulation of supergravity imposes the kinematic constraints (torsion conditions) on the supervielbein and the superconnection, which are necessary to reproduce the correct component fields but make these superfields functionally dependent. Introducing prepotentials solves these constraints, much like introducing a vector potential solves the Bianchi identity in electromagnetism. However, for extended supergravities ($\mathcal{N} \geq 2$) the structure of the constraints becomes extremely intricate and their solutions in terms of the unconstrained prepotentials are still unknown for many theories. 

For $\mathcal{N}=1$ supergravity, the prepotential was identified by Ogievetsky and Sokatchev \cite{Ogievetsky:1976qc, Ogievetsky:1978hg, Ogievetsky:1978mt, Ogievetsky:1979bb}. According to the Ogievetsky–Sokatchev geometric approach to the off-shell \(\mathcal{N}=1\) supergravity, the prepotential is given by an unconstrained axial superfield \(H_m(x, \theta, \bar{\theta})\) whose role is to encode the entire supergeometry without imposing the usual covariant torsion constraints by hand. The key idea is to consider the real superspace $\mathbb{R}^{4|4}$ as a \((4,4)\)-dimensional surface embedded in a complex superspace~$\mathbb{C}^{4|2}$ and then use the gauge supergroup that leaves the chiral
superspace invariant to bring this surface to a flat form up to higher orders, with the prepotential parametrizing the deviation from flat superspace in the Wess-Zumino-type gauge. This construction leads to a formulation where the $\mathcal{N}=1$ supergravity multiplet is described by an unconstrained prepotential, which serves as a fundamental building block from which the supervielbeins and the connections are derived, ensuring that all constraints of minimal supergravity are satisfied automatically \cite{Ogievetsky:1979ay, Ogievetsky:1980de, Ogievetsky:1980qm, Ogievetsky:1980qp}. 

In the framework of $\mathcal{N}=2$ supergravity, the prepotential cannot be obtained in a similiar way. Crucially, the 
explicit identification of the unconstrained prepotentials for extended 
supersymmetries only became possible after the breakthrough discovery of harmonic 
superspace~\cite{Galperin:1984mln, Galperin:1984av}. The harmonic 
superspace can be constructed by expanding standard $\mathcal{N}=2$ 
superspace $\mathbb{R}^{4|8}$ with the additional internal bosonic coordinates named harmonics that 
parametrize an $S^2\sim SU(2)/U(1)$ sphere. This extended space allows the
introduction of analytic superfields and the fundamental supergravity gauge prepotentials 
emerging as the unconstrained analytic  connection in the covariant harmonic derivative \cite{Galperin:1987em, Galperin:1987ek}:
\begin{equation*}
\begin{split}
&\mathcal{N}=2\;\text{conformal supergravity:}
\qquad
\mathfrak{D}^{++} =  \partial^{++} + \mathcal{H}^{++m}\partial_m 
+
\mathcal{H}^{++\hat{\alpha}+} \partial^-_{\hat{\alpha}}
 +
 \mathcal{H}^{(+4)} \partial^{--},
\\
&\mathcal{N}=2\;\text{Einstein supergravity:}
\qquad \; \;\;
\mathbb{D}^{++} =  \partial^{++} + H^{++m}\partial_m 
+
H^{++\hat{\alpha}+} \partial^-_{\hat{\alpha}}
 +
 H^{++5} \partial_5.
 \end{split}
\end{equation*}
 By encoding the entire geometric structure and the superspace 
curvatures without requiring restrictive kinematic constraints to be put by hand, the analytic prepotentials
naturally resolve the long-standing problem of providing an exact off-shell 
formulation for the $\mathcal{N}=2$ supergravity multiplet. 

\smallskip

Within the harmonic formulation of supergravity, many interesting results have been obtained. 
These results include the formulation of various versions of Einstein supergravity, including new principal Einstein supergravity
\cite{Galperin:1987em, Galperin:1987ek, Galperin:1985tn, Delamotte:1987yn, Sokatchev:1988aa, 18, Ivanov:2022vwc}, the construction of the general minimal coupling of the $\mathcal{N} =2$ Yang-Mills fields to $\mathcal{N} = 2$ sigma models in curved harmonic superspace \cite{Bagger:1987rc}, the description of quaternionic manifolds, with one-to-one correspondence between quaternionic spaces and the off-shell $\mathcal{N}=2$ supersymmetric sigma-models coupled to $\mathcal{N}=2$ supergravity
\cite{Ivanov:1987ih, Galperin:1992pj, Ivanov:1998de, Ivanov:1998ih, Ivanov:1999vg, Casteill:2001ap, Casteill:2001zk}.
Nevertheless, many issues related to the harmonic superspace formulation of $\mathcal{N}= 2$
supergravity have never been addressed.
One of the most important problems is the relation between analytic prepotentials and the differential geometric structure of the superspace, i.e. with the full algebra of
covariant derivatives.
Another important problem addressed in this paper is a detailed investigation of the Wess–Zumino-type gauge and the component reduction of harmonic superfield actions on a supergravity background. 

\smallskip
It is worth noting that the procedure of component reduction for superfield theories of supergravity may be quite involved and challenging technically. 
Different methods have been actively developed and applied.
In $\mathcal{N}=1$ superfield supergravity, there are three methods of component reduction:
\begin{enumerate}
	
	\item  \textit{Wess-Zumino gauge on supergravity prepotentials} \cite{BK}
	
	This method is available only when the formulation in terms of unconstrained prepotentials is known. In such a formulation, all the covariant derivatives are constructed from the prepotentials describing the Weyl multiplet and the compensators. After imposing Wess–Zumino gauge on them, the component reduction symplifies to Grassmann integration.
	
	\item \textit{Wess-Zumino gauge on superfield veilbeins and connection} \cite{Wess:1978ns, Muller:1982gn, Gates:1983nr, Wess:1992cp}
	
	In this approach, the fields of the supergravity multiplet are identified with the lowest components of the vielbein in Wess-Zumino gauge and the torsions. Then all the components of the torsion and the curvature supertensors can be constructed in terms of these fields. Also, chiral densities in this approach are constructed by a Noether-type procedure based on the known local supersymmetry transformations.
	
	\item \textit{Superspace normal coordinates} \cite{Gates:1997ag, Grisaru:1997ub}
	
	This is a universal geometric framework for systematically reducing superfield actions to their component field forms. The core idea is to construct a local coordinate system in the curved superspace \cite{McArthur:1983fm, Atick:1986jr}, analogous to Riemannian normal coordinates in gravity, but expanded around the bosonic submanifold (the ordinary spacetime) along the fermionic directions. This covariant approach circumvents the traditional difficulties associated with expanding the superdeterminant of the supervielbein, as it yields the equations that allow one to reconstruct the integration measure without any need for the full, intricate details of the supervielbein. 
	
\end{enumerate}

\smallskip

The problem of component reduction in 
$\mathcal{N}=2$ superspace supergravity has also been actively investigated within the covariant approach, in which the unconstrained supergravity prepotentials remained \textit{terra incognita}.
Of particular note is Ref. \cite{Kuzenko:2008ry}, where two methods were employed to reduce the $\mathcal{N}=2$ projective superspace actions to components, developed in \cite{Kuzenko:2008ep, Kuzenko:2008qw}: the first one based on the principle of the projective invariance, and the other on the normal-coordinate method in curved superspace.
Then, the developed
projective superspace component reduction rule has been applied to the vector-tensor multiplets \cite{Butter:2012xg}.
Interesting results were also obtained in Ref.~\cite{Butter:2015nza}, where a fully covariant approach to harmonic superspace, i.e. the one based on the conformal superspace description of $\mathcal{N}=2$ conformal supergravity~\cite{Butter:2011sr}, was developed.  On the basis of this formalism, a method of the covariant component reduction was proposed.
However, to the best of our knowledge, practical application of this method is non-trivial because of the problem of harmonic singularities. In particular, an analytic measure for AdS$_4$
was constructed in Ref.~\cite{Ivanov:2025jdp}, while Ref.~\cite{Gargett:2025xcg} proposed an alternative analytic action principle designed for the superspace description and component reduction of the free AdS$_4$
hypermultiplet.

\smallskip

In this work, we present a comprehensive analysis of Wess–Zumino gauge structure for the $\mathcal{N}=2$ supergravity analytic prepotentials, which, to date, has not yet been fully explored\footnote{A number of results were obtained in the linearized approximation; see Refs.~\cite{Zupnik:1998td, Buchbinder:2021ite, Buchbinder:2025ceg, Ivanov:2024gjo, Ivanov:2025mld}.}. We consider this to be important for the following reasons:
\begin{enumerate}
		\item  
		These results provide a basis for studying the component structure of the harmonic superfield theories on the $\mathcal{N}=2$ supergravity background. 
		The method applied is the full analogue of the prepotential method of the component reduction in 
		$\mathcal{N}=1$ supergravity. 
	
	\item 
		Crucially, in the case of $\mathcal{N}=2$ the harmonic approach is based on the analyticity principle, which plays a fundamental role in the formulation of all known
		$\mathcal{N}=2$ superfield theories. 
		In particular, the $\mathcal{N}=2$ supergravity prepotentials are unconstrained analytic superfields, which marks the analytic basis as the preferable one.
		  The earlier works on component analysis suggested the central basis, thereby preventing this fundamental principle from being manifest.
		One of the goals of the present paper is to fill this gap and demonstrate how component reduction works in the analytic basis.
		
		\item Moreover, these results are useful for the full harmonic description of curved $\mathcal{N}=2$ superspaces, in particular, $\mathcal{N}=2$ AdS superspace. The AdS superspace is of great interest since it is necessary for constructing the harmonic formulation of the linearized $\mathcal{N}=2$ AdS supergravity and $\mathcal{N}=2$  AdS higher-spin theories which generalize those known in 4D flat space \cite{Zupnik:1998td, Buchbinder:2021ite} (see \cite{Buchbinder:2025ceg, Buchbinder:2026nwu} for a recent review ).
		In view of the fundamental role of AdS superspace in higher-spin theory \cite{Fradkin:1987ks, Vasiliev:1990en}, this problem is of primary importance in the context of 
		$\mathcal{N}=2$ higher-spin theories. 
\end{enumerate}	

\smallskip

We analyze both the $\mathcal{N}=2$ conformal supergravity multiplet and the $\mathcal{N}=2$  Einstein supergravity multiplet with the nonlinear multiplet compensator (the version in components was first discovered by Fradkin, Vasiliev and by de Wit, van Holten, van Proeyen \cite{Fradkin:1979cw, Fradkin:1979as, deWit:1979dzm}). The main goal of this paper is to perform  a detailed study of Wess-Zumino gauge in the \textit{bosonic sector}, i.e. to obtain the complete expressions for the field redefinitions and derive their full nonlinear transformation laws. We provide a detailed systematic description of all the steps of this procedure.  As an application of the obtained results, we consider the derivation of the bosonic sector of the component action for the hypermultiplet in the off-shell $\mathcal{N}=2$ supergravity background. The results for the fermionic and the mixed sectors, along with some further applications, we plan to present in a separate publication.

\smallskip

This paper is organized as follows. 
In section \ref{sec: N=2 superfield supergravity}, we discuss the superfield prepotentials of the $\mathcal{N}=2$ Weyl and Einstein supergravity multiplets, their gauge transformations, and their interactions with the hypermultiplet. Section \ref{sec: conf sg} is devoted to Wess—Zumino gauge for the $\mathcal{N}=2$ Weyl supermultiplet and the component content of their hypermultuplet coupling. In Section \ref{sec: Ein sg}, we perform the similar analysis for the $\mathcal{N}=2$ Einstein supergravity multiplet.
The last section~\ref{sec: dis}
summarizes the obtained results and presents some related problems. We also include several appendices. In appendix \ref{app: notations}, we provide the summary of notations and conventions. In
Appendix \ref{eq: gauge SU(2)}  we discuss the technical details of the $SU(2)$ gauge transformations.

\section{$\mathcal{N}=2$ supergravity theories in harmonic superspace}\label{sec: N=2 superfield supergravity}

In this section, we present the superfield formulation of the Weyl and Einstein $\mathcal{N}=2$ supergravity multiplets in harmonic superspace, as well as the superfield action of the hypermultiplet on the supergravity background. It is precisely these models that we investigate in this paper. In this section, we follow Refs. \cite{18, Ivanov:2022vwc, Galperin:1987ek, Galperin:1987em, Galperin:1985zv}.

\subsection{Gauge group of $\mathcal{N}=2$ conformal supergravity}

The analytic action that describes the coupling of the
$\mathcal{N}=2$ Weyl supermultiplet to the hypermultiplet reads
	\begin{equation}\label{eq: action}
		S_{hyp} [q, \mathcal{H}] = - \frac{1}{2} \int d\zeta^{(-4)}\, q^{+a} \mathfrak{D}^{++}q^+_a,
	\end{equation}	
where $q^{+a} = (\tilde{q}^+, q^+)$, $q^+_a = \epsilon_{ab} q^{+b} = (q^+, -\tilde{q}^+)$  form Pauli-G\"ursey hypermultiplet doublet. The operator $\mathfrak{D}^{++}$ denotes the covariant harmonic derivative,
\begin{equation}\label{eq: cov der}
    \mathfrak{D}^{++} := \partial^{++} 
    +
    \mathcal{H}^{++m}(\zeta) \partial_m 
    +
    \mathcal{H}^{++\hat{\alpha}+} (\zeta)\partial^-_{\hat{\alpha}} 
     +
    \mathcal{H}^{++\hat{\alpha}-} (\zeta, \theta^-)\partial^+_{\hat{\alpha}} 
    +
    \mathcal{H}^{(+4)}(\zeta) \partial^{--},
\end{equation}
where the analytic superfields $\mathcal{H}^{++}$ describe the off-shell $\mathcal{N}=2$ Weyl multiplet.
As the hypermultiplet superfield $q^+_a$ satisfies the Grassmann analyticity condition $\partial^+_{\hat{\alpha}} q^+_a = 0$, a non-analytic superfield $\mathcal{H}^{++\hat{\alpha}-}$ does not enter the action \eqref{eq: action}.

\smallskip

The covariant harmonic derivative satisfies the reality condition with respect to harmonic conjugation, see Appendix \ref{app: notations}:
\begin{equation}
	\widetilde{\mathfrak{D}^{++}} 
	=
	\mathfrak{D}^{++}
	\qquad
	\Rightarrow
	\qquad
	\begin{cases}
		\widetilde{ \mathcal{H}^{++m}} =  \mathcal{H}^{++m},
		\\
		\widetilde{  \mathcal{H}^{++\alpha\pm}}
		=
		  \mathcal{H}^{++\dot{\alpha}\pm},
		  \\
		  \widetilde{ \mathcal{H}^{++\dot{\alpha}\pm}}
		  =
		  - \mathcal{H}^{++\alpha\pm},
		\\
		\widetilde{\mathcal{H}^{(+4)}}
		=
		  \mathcal{H}^{(+4)}.
	\end{cases}	
\end{equation}	
This leads to the standard reality conditions on the fields in the component expansion, which we apply in subsequent sections.

\smallskip

The hypermultiplet analytic action \eqref{eq: action} constitutes one of the contributions to the action of the principal version of $\mathcal{N}=2$ Einstein supergravity \cite{Galperin:1987ek}. Furthermore, this action provides a clear illustration of the realization of the $\mathcal{N}=2$ supergravity gauge group and explains the natural geometric origin of the prepotentials as the supervielbeins in the harmonic derivative (see also discussion in \cite{Buchbinder:2025ceg}).

\smallskip

The hypermultiplet action \eqref{eq: action}	is invariant under the analyticity-preserving general coordinate transformations:	
	\begin{equation}\label{eq: coord transf}
			\begin{split}
				&x^{m} \,\,\to x^{\prime m} = x^{m} + \lambda^{m}(\zeta),
				\\
				&\theta^{+\hat{\alpha}} \to  	\theta^{\prime+\hat{\alpha}} = 	\theta^{+\hat{\alpha}} + 	\lambda^{+\hat{\alpha}}(\zeta),
				\\
				&\theta^{-\hat{\alpha}} \to  	\theta^{\prime-\hat{\alpha}} = 	\theta^{-\hat{\alpha}} + 	\lambda^{-\hat{\alpha}}(\zeta, \theta^-),
				\\
				&
				u^{+i} \,\to u^{\prime +i} = u^{+i} + \lambda^{++}(\zeta) u^{-i},
				\\
				&
				u^{-i} \,\to u^{\prime -i} = u^{-i},
			\end{split}
	\end{equation}	
which act on the superfields according to the rules\footnote{The harmonic derivative $\mathcal{D}^0$ is defined as 
$
\mathcal{D}^0 : = \partial^0 + \theta^{+\hat{\alpha}} \partial^-_{\hat{\alpha}} 
-
\theta^{-\hat{\alpha}} \partial^+_{\hat{\alpha}}
$. Its action on any harmonic superfield gives its harmonic charge $\mathcal{D}^0 \Phi^{(q)} = q \Phi^{(q)} $.}
    	\begin{equation}\label{eq: gauge transformations}
		\begin{split}
			\mathfrak{D}^{++} (\zeta, \theta^-)\; &\to\; \mathfrak{D}^{\prime ++} (\zeta^\prime, {\theta^-}^\prime) = 
			\mathfrak{D}^{++} (\zeta, \theta^-) - \lambda^{++}(\zeta) \mathcal{D}^0,
			\\
			 q_a^{+}(\zeta)	
			&\to
			q_a^{\prime+}(\zeta^\prime)	
			=
			\left(\text{Ber} \left|\frac{\partial \zeta }{\partial \zeta^\prime}  \right| \right)^{\frac{1}{2}} q_a^{+}(\zeta).
		\end{split}
	\end{equation}	
These transformations constitute the correct gauge group for $\mathcal{N}=2$ conformal supergravity \cite{Galperin:1984av, 18, Ivanov:2022vwc}. In this paper, we demonstrate this in detail in the bosonic sector.
We emphasize that, unlike other superfield formulations of $\mathcal{N}=2$ supergravity (see, e.g., \cite{Kuzenko:2022ajd}), in the harmonic formulation the Lorentz (super)transformations are encoded within the world-diffeomorphism supergroup. This is fully analogous to the $\mathcal{N}=1$ supergravity formulation by Ogievetsky and Sokatchev \cite{Ogievetsky:1979bb}.

\smallskip

The transformation law for the covariant harmonic derivative \eqref{eq: gauge transformations} is motivated by the realization of the rigid $\mathcal{N}=2$ superconformal group $SU(2,2|2)$ in the harmonic superspace \cite{Galperin:1985zv, 18, Buchbinder:2024pjm, Ivanov:2025gvs}.
To clarify the appearance of the $\lambda^{++} \mathcal{D}^0$ term on the right-hand side (RHS) of \eqref{eq: gauge transformations}, let us restrict to the transformations of the harmonics parametrized by $\lambda^{++}$ and consider the transformation of the partial harmonic derivative $\partial^{++}$:
\begin{equation*}
    \partial^{++} = u^{+i} \frac{\partial}{\partial u^{-i}}
    \quad
    \to
    \quad
    \partial^{\prime++} = u^{\prime+i} \frac{\partial}{\partial u^{\prime-i}}
    =
    (u^{+i} + \lambda^{++}(\zeta) u^{-i})
    \left[ \frac{\partial u^{-j}}{\partial u^{\prime-i}} \frac{\partial}{\partial u^{-j}} + \frac{\partial u^{+j}}{\partial u^{\prime-i}} \frac{\partial}{\partial u^{+j}} \right].
\end{equation*}
Using relations $u^{\prime-i} = u^{-i}  $ and $u^{+i} = u^{\prime +i} - \lambda^{++} u^{\prime -i}$, we obtain
\begin{equation*}
    \partial^{\prime ++} = \partial^{ ++}
    -
    \lambda^{++}  \underbrace{\left( u^{+i} \frac{\partial}{\partial u^{+i}}   - u^{-i} \frac{\partial}{\partial u^{-i}}  \right)}_{\partial^0}
    -
    (\partial^{\prime ++} \lambda^{++}) \partial^{--}.
\end{equation*}
Thus the derivative $\partial^0$
  naturally appears as a result of the harmonic transformation.
 In principle, one could incorporate an appropriate $\partial^0$-vielbein into the covariant derivative \eqref{eq: cov der}. However, this vielbein is redundant as it can be eliminated by redefinition of $\mathcal{H}^{++\hat{\alpha}\pm}$ (recall that $\partial^0 = \mathcal{D}^0 - \theta^{+\hat{\alpha}} \partial^-_{\hat{\alpha}} - \theta^{-\hat{\alpha}} \partial^+_{\hat{\alpha}}$ and $q^{+a}\mathcal{D}^0 q^{+}_a =0$). This also explains why no extra vielbein is needed for the $\partial^{++}$-derivative: its coefficient remains unchanged under the coordinate transformations~\eqref{eq: coord transf}.

\smallskip

The invariance of the hypermultiplet action \eqref{eq: action} under transformations \eqref{eq: coord transf} and \eqref{eq: gauge transformations} can be verified in a straightforward manner:
    \begin{equation}
    \begin{split}
	S_{hyp}^\prime[q^\prime, \mathcal{H}^\prime] &= \frac{1}{2} \int d\zeta^{\prime(-4)}\, q^{\prime+a} \mathfrak{D}^{\prime++}q^{\prime+}_a
        \\&=
        \frac{1}{2} \int d\zeta^{(-4)} \text{Ber} \left|\frac{\partial \zeta^\prime }{\partial \zeta}  \right|\, \left(\text{Ber} \left|\frac{\partial \zeta }{\partial \zeta^\prime}  \right| \right)^{\frac{1}{2}} q^{+a} \left(\mathfrak{D}^{++} - \lambda^{++} \mathcal{D}^0\right) \left\{ \left(\text{Ber} \left|\frac{\partial \zeta }{\partial \zeta^\prime}  \right| \right)^{\frac{1}{2}} q^{+}_a \right\}
        .
        \end{split}
	\end{equation}	
After applying the property $q^{+a} q^+_a = 0$, many terms cancel, and we obtain:
\begin{equation}
    S_{hyp}^\prime[q^\prime, \mathcal{H}^\prime]
    =
     \frac{1}{2} \int d\zeta^{(-4)}  q^{+a} \mathfrak{D}^{++}   q^{+}_a 
      =
      S_{hyp} [q, \mathcal{H}].
\end{equation}

The full action for $\mathcal{N}=2$ conformal supergravity (generalizing the square of the Weyl tensor) in terms of the fundamental analytic $\mathcal{H}^{++}$ prepotentials is currently unknown. Nevertheless, there are some conjectures regarding its structure, see the recent work \cite{Ivanov:2025gvs}.

\subsection{Gauge group of $\mathcal{N}=2$ Einstein supergravity}\label{eq: sec Einstein sg gg}

One can also consider an alternative group of analyticity-preserving  coordinate transformations:
\begin{equation}\label{eq: coord transf Einstein}
			\begin{split}
				&x^{m} \,\,\to x^{\prime m} = x^{m} + \lambda^{m}(\zeta),
				\\
				&\theta^{+\hat{\alpha}} \to  	\theta^{\prime+\hat{\alpha}} = 	\theta^{+\hat{\alpha}} + 	\lambda^{+\hat{\alpha}}(\zeta),
				\\
				&\theta^{-\hat{\alpha}} \to  	\theta^{\prime-\hat{\alpha}} = 	\theta^{-\hat{\alpha}} + 	\lambda^{-\hat{\alpha}}(\zeta, \theta^-),
				\\
				&
				u^{\pm i} \,\to u^{\prime \pm i} = u^{\pm i}.
			\end{split}
	\end{equation}	
In contrast to the transformations \eqref{eq: coord transf} of $\mathcal{N}=2$ conformal supergravity, this set leaves the harmonics $u^{\pm i}$ completely inert. These transformations constitute the gauge group of the $\mathcal{N}=2$ Einstein supergravity, version with the non-linear tensor compensator  \cite{Galperin:1987em,Galperin:1987ek,  Ivanov:2022vwc, 18}, whose component formulation was discovered in \cite{Fradkin:1979cw, Fradkin:1979as} and \cite{deWit:1979dzm}. 

Within the standard compensator approach, this group arises from the $\mathcal{N}=2$ conformal supergravity group \eqref{eq: coord transf} after imposing appropriate gauge conditions on the conformal compensators (or, equivalently, by eliminating the redundant gauge degrees of freedom associated with the harmonic $\lambda^{++}$ transformations). 
For this purpose, we should take the non-linear multiplet as a compensator.
\smallskip 

\noindent  \textbf{Non-linear multiplet}

\smallskip

On the $\mathcal{N}=2$ conformal supergravity background, the non-linear multiplet is represented by an analytic superfield $N^{++}(\zeta)$ of harmonic charge $+2$ which satisfies the constraint
\begin{equation}\label{eq: NL mult constr}
    \mathfrak{D}^{++} N^{++} + (N^{++})^2 - \mathcal{H}^{(+4)} = 0.
\end{equation}
This constraint is invariant under transformations \eqref{eq: gauge transformations} provided that $N^{++}$ transforms as
\begin{equation}
    \delta_\lambda N^{++}= \lambda^{++}.
\end{equation}
We can now fix the gauge by setting $N^{++}=0$. Alongside the transformation law, this condition forces $\lambda^{++}=0$. Substituting $N^{++}=0$ into constraint \eqref{eq: NL mult constr} yields $\mathcal{H}^{(+4)}=0$. Consequently, the $\mathcal{N}=2$ conformal superdiffeomorphism group \eqref{eq: coord transf} reduces to the smaller group~\eqref{eq: coord transf Einstein} under whose action the harmonics $u^{\pm i}$ are inert.

\newpage

\noindent\textbf{$\mathcal{N}=2$ vector multiplet} 

\medskip
A single compensator is not sufficient for the formulation of $\mathcal{N}=2$ Einstein supergravity (e.g. see the discussion in \cite{18}).
The second conformal compensator is realised as the $\mathcal{N}=2$ vector multiplet. 
The vector multiplet can be described as an analytic supervielbein $H^{++5}$ for an auxiliary coordinate $x^5$
 associated with the central charge. The corresponding covariant harmonic derivative takes the form\footnote{We deliberately use different notations for the analytic prepotentials and the covariant harmonic derivative to distinguish the conformal case from the Einstein one.
 }
\begin{equation}\label{eq: harm der EI}
    \mathbb{D}^{++} = \partial^{++} + H^{++m}(\zeta) \partial_m 
    +
    H^{++\hat{\alpha}+} (\zeta)\partial^-_{\hat{\alpha}} 
     +
    H^{++\hat{\alpha}-} (\zeta, \theta^-)\partial^+_{\hat{\alpha}} 
    +
    H^{++5}(\zeta) \partial_5.
\end{equation}
Gauge transformations of the $H^{++5}$ prepotential are generated by analytic shifts of the auxiliary coordinate $x^5$:
\begin{equation}
    	x^5 \;\;\to x^{\prime 5} = x^5 + \lambda^5(\zeta),
\end{equation}
under which $\mathbb{D}^{++}$ remains invariant:
\begin{equation}
    \mathbb{D}^{++}(\zeta, \theta^-, x^5) \;\to\; \mathbb{D}^{\prime ++}(\zeta^\prime, {\theta^-}^\prime, x^5) = \mathbb{D}^{++}(\zeta, \theta^-, x^5).
\end{equation}

The action of the hypermultiplet on the Einstein supergravity background has the same form as \eqref{eq: action}, with the replacement $\mathfrak{D}^{++} \to \mathbb{D}^{++}$.  
For a massive hypermultiplet carrying a nontrivial central charge, the superfields acquire the dependence on $x^5$:
\begin{equation}
    \mathbf{q}^+(\zeta, x^5) = e^{i m_q x^5} q^+(\zeta), \qquad
    \mathbf{\tilde{q}}^+(\zeta, x^5) = e^{-i m_q x^5} \tilde{q}^+(\zeta).
\end{equation}

For completeness, we present here the action for the $\mathcal{N}=2$ Einstein supergravity case:
\begin{equation}\label{eq: vect mult}
    S_{Ein} = - \frac{1}{4\kappa^2} \int dx^4 d^8\theta du\,  E  H^{++5} H^{--5}.
\end{equation}
Here $E$ denotes the measure of the full harmonic superspace and can be constructed in terms of analytic prepotentials, while the superfield $H^{--5}$ is defined as a solution of the covariantized zero-curvature equations
\begin{equation}
    [\mathbb{D}^{++}, \mathbb{D}^{--}] = \mathcal{D}^0,
\end{equation}
where the negatively charged harmonic derivative is given by
\begin{equation}\label{eq: harm der EI -}
	\mathbb{D}^{--} := \partial^{--} + H^{--m}(\zeta) \partial_m 
	+
	H^{--\hat{\alpha}+} (\zeta)\partial^-_{\hat{\alpha}} 
	+
	H^{--\hat{\alpha}-} (\zeta, \theta^-)\partial^+_{\hat{\alpha}} 
	+
	H^{--5}(\zeta) \partial_5.
\end{equation}

\section{$\mathcal{N}=2$ conformal supergravity}\label{sec: conf sg}

Now we turn to the analysis of the Wess–Zumino-type gauge for the $\mathcal{N}=2$ conformal supergravity multiplet. Even for the simplest case of $\mathcal{N}=1$ supergravity, a complete study of the component structure of the supergravity prepotentials and the component fields' nonlinear transformations is a nontrivial, time-consuming task (see Ref. \cite{Ogievetsky:1979bb, BK}). In particular, a correct matching of components requires field redefinitions. 
The aim of this section is to investigate the Wess—Zumino gauge for the bosonic sector of $\mathcal{N}=2$ conformal supergravity in full detail, deriving the non-linear transformation laws for the component fields and determining necessary field redefinitions.

Recall that the dynamical degrees of freedom of the off-shell $\mathcal{N}=2$ conformal supergravity are described by the superfields within the covariant harmonic derivative \eqref{eq: cov der}.
Under infinitesimal gauge transformations \eqref{eq: gauge transformations}, $\mathfrak{D}^{++}$ transforms as\footnote{By definition, for every superfield one has
\begin{equation*}
    \delta_\lambda \Phi (x,\theta, u^\pm)
    : = 
    \Phi^\prime (x^\prime,\theta^\prime, u^{\pm\prime})
    -
    \Phi (x,\theta, u^\pm)
    =
    \Phi^\prime (x,\theta, u^{\pm})
    +
    \lambda^M \partial_M \Phi (x,\theta, u^\pm) 
    -
    \Phi (x,\theta, u^\pm).
\end{equation*}
We adopt the convention that $\delta_\lambda$ denotes \textit{active} transformations. The
\textit{passive} ones, defined by $\delta^*_\lambda \Phi := \Phi'(\zeta) - \Phi(\zeta)$, are then related to them by translational terms: $\delta^*_\lambda \Phi = \delta_\lambda \Phi - \lambda^M \partial_M \Phi$. In what follows, we will work with passive transformations when studying the component structure.
}
\begin{equation}\label{eq: spin 2 ct}
	\delta_\lambda \mathfrak{D}^{++}
    =
    \mathfrak{D}^{\prime ++}(\zeta^\prime, \theta^{\prime-})
    -
    \mathfrak{D}^{++}(\zeta, \theta^-)
    = 
    - \lambda^{++} \mathcal{D}^0.
\end{equation}	
This condition determines the infinitesimal transformation laws of the $\mathcal{H}^{++}$ prepotentials:
\begin{equation}\label{eq: H transf}
	\begin{split}
		\delta_\lambda \mathcal{H}^{++m} &= \mathfrak{D}^{++}  \lambda^{m},
		\\ 
		\delta_\lambda \mathcal{H}^{++\hat{\alpha}+} &= \mathfrak{D}^{++} \lambda^{+\hat{\alpha}} 
		-
		\lambda^{++} \theta^{+\hat{\alpha}},
		\\
		\delta_\lambda \mathcal{H}^{++\hat{\alpha}-} &= \mathfrak{D}^{++} \lambda^{-\hat{\alpha}} + \lambda^{++} \theta^{-\hat{\alpha}},
		\\
		\delta_\lambda 
		\mathcal{H}^{(+4)} &= \mathfrak{D}^{++} \lambda^{++}.
	\end{split}	
\end{equation}	
Using the non-analytic gauge parameter $\lambda^{-\hat{\alpha}}(\zeta,\theta^-)$, one can impose the \textit{analytic gauge} 
\begin{equation}
 \mathcal{H}^{++\hat{\alpha}-} = \theta^{+\hat{\alpha}}
 \quad
 \Rightarrow
 \quad
 \mathfrak{D}^{++} \lambda^{-\hat{\alpha}} + \lambda^{++} \theta^{-\hat{\alpha}} = \lambda^{+\hat{\alpha}}.
 \end{equation}
It means that the  non-analytic superfield $\mathcal{H}^{++\hat{\alpha}-}$ contains only pure gauge degrees of freedom. In what follows we work exclusively in the analytic gauge, so that the harmonic derivative $\mathfrak{D}^{++}$ satisfies the analyticity condition $[\mathcal{D}^+_{\hat{\alpha}}, \mathfrak{D}^{++}] = 0$. 

\smallskip

We can impose the Wess—Zumino-type gauge by applying the gauge transformations \eqref{eq: H transf}.
Its main purpose is to isolate the physical degrees of freedom and eliminate the unphysical (redundant) ones. Moreover, it removes non-polynomial terms that would otherwise appear in interacting models. In this paper, we study the \textbf{bosonic sector} of the $\mathcal{N}=2$ Weyl multiplet in detail.

\medskip 

\subsection{Wess–Zumino-type gauge}

To this end, we start with the general expansions of the analytic prepotentials in the bosonic sector:
\begin{subequations}
\begin{equation}\label{eq: H fields general}
\begin{split}
    \mathcal{H}^{++n} = \;&L^{++n}(x, u)  + (\theta^+)^2 C^n(x,u)  +(\bar{\theta}^+)^2 \tilde{C}^n(x, u) 
    \\&
    \qquad  \qquad  \qquad  \qquad  \qquad + i \theta^{+} \sigma^a \bar{\theta}^+ M^n_a(x, u)  + (\theta^+)^4 K^{--n}(x, u) ,
\\
    \mathcal{H}^{++\hat{\beta}+} = \;& \theta^{+ \alpha} A^{+\hat{\beta}+ }_{\alpha}(x, u) + \bar{\theta}^{+ {\dot{\alpha}}} B^{+\hat{\beta}+}_{\dot{\alpha}} (x, u) + (\theta^+)^2 \bar{\theta}^{+\dot{\alpha}} P^{\hat{\beta}}_{\dot{\alpha}}(x,u) + (\bar{\theta}^+)^2 \theta^{+\alpha} T_\alpha^{\hat{\beta}}(x, u) ,
\\
    \mathcal{H}^{(+4)} = \;&L^{(+4)}(x, u) + (\theta^+)^2 C^{(+2)}(x,u)  +(\bar{\theta}^+)^2 \tilde{C}^{(+2)}(x, u) 
     \\&
    \qquad  \qquad  \qquad  \qquad  \qquad  + i \theta^{+} \sigma^a \bar{\theta}^{+} M^{(+2)}_a(x, u)  
     + (\theta^+)^4 K(x, u) 
\end{split}
\end{equation}
and define the component expansions for the analytic supergauge parameters as follows: 
\begin{equation}\label{eq: lambda par general}
\begin{split}
    \lambda^n =\;&l^n(x, u) + (\theta^+)^2 c^{(-2)n}(x,u)  +(\bar{\theta}^+)^2 \tilde{c}^{(-2)n}(x, u)  \\&
    \qquad  \qquad  \qquad  \qquad  \qquad  + i \theta^{+} \sigma^a \bar{\theta}^{+} m^{(-2)n}_a(x, u) + 
     (\theta^+)^4 k^{(-4)n}(x, u), 
     \\
    \lambda^{+\hat{\beta}} =&\; \theta^{+ \alpha} a^{\hat{\beta}}_{\alpha}(x, u) + \bar{\theta}^{+{\dot{\alpha}}} \bar{b}^{\hat{\beta}}_{\dot{\alpha}} (x, u) + (\theta^+)^2 \bar{\theta}^{+\dot{\alpha}} \bar{g}^{(-2)\hat{\beta}}_{\dot{\alpha}}(x,u) + (\bar{\theta}^+)^2 \theta^{+\alpha} j_\alpha^{(-2)\hat{\beta}}(x, u) ,
    \\
    \lambda^{++} =\; &l^{++}(x, u) + (\theta^+)^2 c(x,u)  +(\bar{\theta}^+)^2 \tilde{c}(x, u) + i \theta^{+} \sigma^a \bar{\theta}^{+} m_a(x, u)  + (\theta^+)^4 k^{(-2)}(x, u).
    \end{split}
\end{equation}
\end{subequations}
All components of the expansions are arbitrary functions of the coordinate $x^m$ and the harmonics $u^{\pm}_i$.
These expansions demonstrate an unusual feature of the formulation of 
$\mathcal{N}=2$ gauge theories in harmonic superspace: the number of pure gauge degrees of freedom and that of gauge parameters are both infinite.

\smallskip

Since gauge transformations \eqref{eq: H transf} are nonlinear and have a rather complex structure, it is convenient to carry out the analysis of gauge freedom in few stages:

\begin{enumerate}
    \item
   First, we consider only the partial harmonic derivative $\partial^{++}$ term in the gauge transformations.
  This allows us to gauge out an infinite set of degrees of freedom and leaves a finite number of non-zero fields that cannot be eliminated this way. 
  This greatly simplifies future analysis.
    
    \item For this finite set of non-zero components, we find the complete nonlinear gauge transformations. In particular, we impose further gauge conditions for the fields with "shift" transformation laws.

    \item Finally, we study the transformation properties of the remaining components and identify the required redefinitions of both the fields and gauge superparameters. The consistency condition of the Wess–Zumino gauge form to be preserved must be carefully checked.
\end{enumerate}
We now proceed to implement this plan step by step.

\smallskip

\noindent \textbf{1.} In the zeroth order approximation, we  consider only the contributions from the partial harmonic derivative $\partial^{++}$:
\begin{subequations}
\begin{equation}
\begin{split}
    \delta^{(0)} \mathcal{H}^{++m} &= \partial^{++} \lambda^m,
    \\
    \delta^{(0)} \mathcal{H}^{++\alpha+} &= \partial^{++} \lambda^{+\alpha},
    \\
     \delta^{(0)} \mathcal{H}^{++\dot{\alpha}+} &= \partial^{++} \bar{\lambda}^{+\dot{\alpha}},
     \\
     \delta^{(0)} \mathcal{H}^{(+4)} &= \partial^{++} \lambda^{++}.
     \end{split}
\end{equation}
Using expansions \eqref{eq: H fields general} and \eqref{eq: lambda par general}, we match the component fields on both sides. 
First, one can observe that the following components of the superparameters are annihilated by $\partial^{++}$:
\begin{equation}\label{lam_res}
\begin{split}
    &\lambda^n_{(0)} = a^n(x),
\\
    &\lambda^{+\beta}_{(0)} = \theta^{+\beta} l(x) + \theta^{+\alpha} l^{(\beta}_{\;\alpha)}(x) + \bar{\theta}^{+\dot{\alpha}} n^\beta_{\dot\alpha}(x) , 
\\
    &\bar{\lambda}_{(0)}^{+\dot{\beta}} = 
    \bar{\theta}^{+\dot{\beta}} \bar{l}(x)
    +
    \bar{\theta}^{+\dot\alpha} \bar{l}^{(\dot{\beta}}_{\;\dot{\alpha})}(x) + \theta^{+\alpha} 
    \bar{n}^{\dot\beta}_{\alpha}(x) ,     
\\    
    &\lambda^{++}_{(0)} = \lambda^{ij}(x)u^+_i u^+_j 
+
(\theta^+)^2 s(x) +(\bar{\theta}^+)^2 \bar{s}(x) + i \theta^{+} \sigma^a\bar{\theta}^{+} k_a(x). 
\end{split}
\end{equation}
The components of $\lambda^M$ that are not annihilated by the harmonic derivative $\partial^{++}$ can be used to set to zero all the prepotential components in the harmonic expansion except for a finite number of fields\footnote{Already at this stage, in the Wess–Zumino gauge, there is no room left for the gauge field $f_m^a$ of local special conformal transformations (parameter 
	$k^a$). This means that in harmonic approach the constraint $e^n_b R_{mn}(M^{ab}) = 0$ is automatically solved.} 
\begin{equation}\label{eq: gauge 0}
\begin{split}
    &\mathcal{H}^{++n}_{(0)} = (\theta^+)^2 C^n(x)  +(\bar{\theta}^+)^2 \bar{C}^n(x) - 2i \theta^{+} \sigma^a \bar{\theta}^{+} e^n_a(x) + 
    (\theta^+)^2 (\bar{\theta}^+)^2 V^{n\,ij}(x) u^-_i u^-_j, 
\\
    &\mathcal{H}^{++\beta+}_{(0)} = 
    +(\theta^+)^2 \bar{\theta}^{+\dot{\alpha}} P^{\beta}_{\dot{\alpha}}(x) 
     +
    (\bar{\theta}^+)^2 \theta^{+\beta} T(x) 
    +
    (\bar{\theta}^+)^2 \theta^{+\alpha} T_{(\alpha}^{\;\beta)}(x),
\\
    &\mathcal{H}^{++\dot{\beta}+}_{(0)} = -(\bar{\theta}^+)^2 \theta^{+\alpha} 
    \bar{P}^{\dot{\beta}}_{\alpha}(x) 
    +
    (\theta^+)^2 \bar{\theta}^{+\dot{\beta}} \bar{T}(x)
    +
    (\theta^+)^2 \bar{\theta}^{+\dot{\alpha}} \bar{T}_{(\dot\alpha}^{\;\dot{\beta})}(x),
\\    
    &\mathcal{H}^{(+4)}_{(0)} = (\theta^+)^2 (\bar{\theta}^+)^2 D(x).
    \end{split}
\end{equation}
\end{subequations}
It is worth noting that superfields $\mathcal{H}^{++n}, \mathcal{H}^{(+4)}, \lambda^n, \lambda^{++}$ are real with respect to harmonic tilde-conjugation \eqref{eq: tilde conj}. This leads to corresponding reality conditions for the fields and gauge parameters in the expansion,
e.g.:
\begin{equation}
	\overline{e_a^m}= e_a^m,
	\qquad
	\overline{V^{m(ij)}} = V^m_{(ij)}.
\end{equation}

\bigskip

\noindent\textbf{2.} The obtained gauge \eqref{eq: gauge 0} can be further simplified.

\medskip

We begin by considering the transformation of field $C^n(x)$ under the action of the parameter $n^a$, which is contained in the fermionic superparameters:
\begin{equation}
    \lambda^{+\beta} \;\to \;\bar{\theta}^{+\dot{\alpha}} n_{\dot{\alpha}}^\beta(x),
    \qquad
    \bar{\lambda}^{+\dot{\beta}} \;\to\; \theta^{+\alpha} \bar{n}_{\alpha}^{\dot{\beta}}(x).
\end{equation}
Using \eqref{eq: H transf}, we obtain the transformation law for $C(x)$:
\begin{equation}
    \delta_n C^m (x) = 
    -
    2i \bar{n}^a e_a^m(x).
\end{equation}
Thus, by using the parameter $n_a$, one can fix $C^m(x) = \bar{C}^m(x) = 0$.

\medskip

Similarly, for the transformation of field $T(x)$ with respect to the parameter $s(x)$ contained in $\lambda^{++}$, we obtain
\begin{equation}
    \delta_s T(x) = - \bar{s}(x).
\end{equation}
We likewise fix the gauge $T(x) = \bar{T}(x)=0$.

\medskip

\noindent\textbf{3.} We are now left with a finite number of bosonic component fields
\begin{subequations}
\begin{equation}\label{eq: H 1}
\begin{split}
    &\mathcal{H}^{++n}_{WZ} =  -2i \theta^{+} \sigma^a \bar{\theta}^{+} e^n_a(x) + 
    (\theta^+)^4 V^{n\,ij}(x) u^-_i u^-_j, 
\\
    &\mathcal{H}^{++\beta+}_{WZ} =
    + (\theta^+)^2 \bar{\theta}^{+\dot{\alpha}} P^{\beta}_{\dot{\alpha}}(x)  
    +
    (\bar{\theta}^+)^2 \theta^{+\alpha} T_{(\alpha}^{\;\beta)}(x),
\\
    &\mathcal{H}^{++\dot{\beta}+}_{WZ} = -(\bar{\theta}^+)^2 \theta^{+\alpha} 
    \bar{P}^{\dot{\beta}}_{\alpha}(x) 
    +
    (\theta^+)^2 \bar{\theta}^{+\dot{\alpha}} \bar{T}_{(\dot\alpha}^{\;\dot{\beta})}(x),
\\    
    &\mathcal{H}^{(+4)}_{WZ} = (\theta^+)^4 D(x)
    \end{split}
\end{equation}
and a finite number of residual gauge parameters
\begin{equation}\label{lam_res_1}
\begin{split}
    &\lambda^n_{(1)} = a^n(x),
\\
    &\lambda^{+\beta}_{(1)} = l(x) \theta^{+\beta}  + l^{(\beta}_{\;\;\alpha)}(x)\theta^{+\alpha} , 
\\
    &\bar{\lambda}_{(1)}^{+\dot{\beta}} = 
    \bar{l}(x) \bar{\theta}^{+\dot{\beta}} 
    +
    \bar{l}^{(\dot{\beta}}_{\;\;\dot{\alpha})}(x) \bar{\theta}^{+\dot\alpha}  ,     
\\    
    &\lambda^{++}_{(1)} = \lambda^{ij}(x)u^+_i u^+_j 
 + i \theta^{+} \sigma^a\bar{\theta}^{+} k_a(x). 
\end{split}
\end{equation}
\end{subequations}
Let us note that under the action of the $\lambda^{+\hat{\beta}}_{(1)}$ transformations, the $\theta$ monomials transform in a homogeneous way, e.g.:
\begin{equation}\label{eq: l exam}
     (\theta^+)^2 \bar{\theta}^{+\dot{\alpha}} 
    \quad \to \quad
     \left[2l(x) + \bar{l}(x)\right] (\theta^+)^2 \bar{\theta}^{+\dot{\alpha}}
     +
\bar{l}^{(\dot{\alpha}}_{\;\;\dot{\beta})}(x) (\theta^+)^2 \bar{\theta}^{+\dot{\beta}}.
\end{equation}
We see that it is natural to identify $\text{Re}\, l(x) = \frac{1}{2}d(x)$ as the local Weyl transformations parameter and $\text{Im}\, l(x) = \frac{1}{2} r(x)$ as that of the local $R$-symmetry ones. 

We now turn to the analysis of the full transformations \eqref{eq: H transf} in the resulting gauge \eqref{eq: H 1},~\eqref{lam_res_1}. For the consistency of this gauge, we must require that no new fields appear after the transformations aside from those already present in \eqref{eq: H 1}, i.e.:
\begin{equation}
    \delta^*_{\lambda, res} \mathcal{H}^{++M}_{WZ} \sim  \mathcal{H}^{++M}_{WZ},
    \qquad
    M = \{m, \alpha+, \dot{\alpha}+, ++ \}.
\end{equation}
This condition both determines the transformations of the component fields and leads to additional corrections to gauge superparameters. It is convenient to consider the effect of the transformations \eqref{lam_res_1} in each of the parameter sectors sequentially.

\subsubsection{$SU(2)$ singlet sector}

\subsection*{Translations -- parameter $a^n(x)$} 

Plugging the parameter $\lambda^m \to a^m(x)$  into $\delta \mathcal{H}_{WZ}^{++m}$, we obtain the following:
\begin{equation}
    \delta_a \mathcal{H}_{WZ}^{++m}  = \mathcal{H}_{WZ}^{++m \prime} + a^n(x) \partial_n \mathcal{H}_{WZ}^{++m} - \mathcal{H}_{WZ}^{++m}  =  \mathcal{H}_{WZ}^{++m} \partial_m a^n(x).
\end{equation}
From here we see that the index $m$ is a
world index.
Then we extract expressions for the infinitesimal variations of the component fields $e^m_a(x)$ and $V^{m(ij)}(x)$:
\begin{equation}
\begin{split}
    &\delta^*_a e^n_a(x)  = e^m_a(x) \partial_m a^n(x) - a^n(x) \partial_n e^m_a(x),
\\
    &\delta^*_a V^{m(ij)}(x) = V^{m(ij)}(x) \partial_m a^n(x) - a^n(x) \partial_n V^{m(ij)}(x) .
    \end{split}
\end{equation}
The first terms correspond to the infinitesimal diffeomorphism
transformation with respect to the world index, while the second ones represent translational contributions.

The remaining component fields in the WZ gauge \eqref{eq: H 1} do not carry a world index, and therefore their transformation laws contain only translational contributions, e.g.:
\begin{equation}
    \delta_a^* T_{(\alpha}^{\;\,\beta)}(x) = - a^n(x)\partial_n T_{(\alpha}^{\;\,\beta)}(x). 
\end{equation}

\subsection*{Weyl and $R$ transformations -- parameter $l(x) = \frac{1}{2}[d(x) + i r(x)]$ }

Let us start by noting that the terms in the transformation law responsible for coordinate changes (see eq. \eqref{eq: l exam}) lead to Weyl transformations for the component fields in $\mathcal{H}_{WZ}^{++m}$ and $\mathcal{H}_{WZ}^{(+4)}$:
\begin{equation}\label{eq: Weyl transf}
    \begin{split}
    &\delta^*_l e_a^{\;n}(x) = -d(x) e_a^{\;n}(x),
    \\
    &\delta^*_l V^{n\,(ij)}(x) = -2 d(x) V^{n\,(ij)}(x), 
    \\
    &\delta^*_l D(x) = -2d(x) D(x).
    \end{split}
\end{equation}
The transformation of the remaining fields is determined by
\begin{equation}
\delta_\lambda \mathcal{H}_{WZ}^{++\hat{\alpha}+} = \mathfrak{D}^{++}_{WZ} \lambda^{+\hat{\alpha}} 
		-
		\lambda^{++} \theta^{+\hat{\alpha}}
\end{equation}  
with derivative terms 
\begin{equation}
    \mathfrak{D}_{WZ}^{++} \lambda^{+\alpha}\Big|_{l(x)}
    =
    i (\theta^+)^2 \bar{\theta}^{+\dot{\beta}} (\sigma^a)^\alpha_{\dot{\beta}} e_a^n(x) \partial_n l(x)
    +
    \left\{  (\theta^+)^2 \bar{\theta}^{+\dot{\alpha}} P^{\beta}_{\dot{\alpha}}(x)  
    +
    (\bar{\theta}^+)^2 \theta^{+\alpha} T_{(\alpha}^{\;\beta)}(x) \right\} l(x).
\end{equation}
Then we obtain:
\begin{equation}\label{eq: trans P1}
\begin{split}
    &\delta^*_l P_a(x) = ie_a^n(x)\partial_n l(x)
    -
    d(x) P_a(x),
    \\
    & \delta^*_l T_{(\alpha}^{\;\beta)}(x) 
    =
    - [d(x)- i r(x)]\,T_{(\alpha}^{\;\beta)}(x).
   \end{split} 
\end{equation}
Field $\text{Im} \, P_a(x)$ serves as the gauge field for local Weyl transformations (parameter $\text{Re}\, l(x) = \frac{1}{2}d(x)$), while $\text{Re}\, P_a(x)$ plays such a role for local
$R$-symmetry (parameter $\text{Im}\, l(x) = \frac{1}{2}r(x)$):
\begin{equation}\label{eq: P transf}
\begin{split}
    &\delta_l \text{Re} P_a(x) = - \frac{1}{2} e_a^n(x) \partial_n r(x) - d(x) \text{Re} P_a(x),
    \\
    & \delta_l \text{Im} P_a(x) = 
    +
    \frac{1}{2} e_a^n(x) \partial_n d(x) - d(x) \text{Im} P_a(x).
\end{split}
\end{equation}

\subsection*{Lorentz transformations -- parameters $l_{(\alpha\beta)}(x)$, $\bar{l}_{(\dot{\alpha}\dot{\beta})}(x)$ }

To begin with, fields $V^{m(ij)}(x)$ and $D(x)$ do not transform under local Lorentz transformations, since they multiply a Lorentz-invariant monomial $(\theta^+)^2 (\bar{\theta}^+)^2$:
\begin{equation}
    \delta^*_{Lor} V^{m(ij)}(x) = 0,
    \qquad
    \delta^*_{Lor} D(x) = 0.
\end{equation}
The infinitesimal transformation of $e^n_a(x)$ is determined by the equation:
\begin{equation}
    \theta^{+\beta} \sigma^a_{\beta \dot{\beta}} \bar{\theta}^{+\dot{\beta}} \delta_{Lor} e^n_a(x)
    +
    l^{(\beta}_{\;\alpha)} \theta^{+\alpha} \sigma^a_{\beta \dot{\beta}} \bar{\theta}^{+\dot{\beta}} e^n_a(x)
    +
    \bar{l}^{(\dot{\beta}}_{\;\dot{\alpha})} \theta^{+\beta} \sigma^a_{\beta \dot{\beta}} \bar{\theta}^{+\dot{\beta}}  e^n_a(x)
    =
    0,
\end{equation}
from which, using eq. \eqref{eq: rel Lor and spin}, we deduce
\begin{equation}
    \delta^*_{Lor} e^n_a(x) = l_{[a}^{\;\;b]}(x) e^n_b(x).
\end{equation}
Thus, the lower index $a$ of field $e^m_a(x)$ is a Lorentz index, and field $e^m_a(x)$ can be naturally identified with the inverse gravitational tetrad.

\smallskip

In the calculation of $\mathcal{H}^{++\alpha+}$ components transformations, the determining contribution is provided by the term
\begin{equation}
    \mathfrak{D}_{WZ}^{++} \lambda^{+\alpha}\Big|_{l_{(\alpha\beta)}}
    =
    -i (\theta^+)^2 \bar{\theta}^{+\dot{\beta}} (\sigma^a)_{\beta\dot{\beta}} e_a^n \partial_n l^{(\alpha\beta)}
    +
     l^{(\alpha}_{\;\;\beta)}
    \left\{  (\theta^+)^2 \bar{\theta}^{+\dot{\alpha}} P^{\beta}_{\dot{\alpha}}  
    +
    (\bar{\theta}^+)^2 \theta^{+\rho} T_{(\rho}^{\;\beta)} \right\}.
    \label{P^a analysis started}
\end{equation}
Using this expression, we obtain:
\begin{equation}
\begin{split}
    &\delta^*_{Lor} P_{\dot{\alpha}}^\alpha = 
    -
    i
    (\sigma^a)_{\beta\dot{\alpha}} e_a^n\partial_n l^{(\alpha\beta)}
    +l^{(\alpha}_{\;\;\beta)} P_{\dot{\alpha}}^\beta
    -
    \bar{l}_{(\dot{\alpha}}^{\;\;\dot{\beta})} P_{\dot{\beta}}^\alpha,
    \\
&\delta^*_{Lor} 
T_{(\alpha\beta)} =- l_{(\alpha}^{\;\;\gamma)} T_{(\gamma\beta)}
- l^{\;\;\gamma)}_{(\beta}T_{(\alpha\gamma)} .
\end{split}
\end{equation}
Upon converting into a Lorentz index and invoking the relations \eqref{eq: rel Lor and spin} and \eqref{eq: D and SD}, we arrive at
\begin{equation}\label{eq: trans P2}
\begin{split}
    \delta^*_{Lor} P^a &=  l^a_{\;\;b} P^b 
    +
    i e_b^n \partial_n (l^{+})^{ab}
    \\
    &=  l^a_{\;\;b} P^b 
      + 
      \frac{i}{2}
    e_b^n \partial_n l^{ab}
    -
    \frac{1}{4}\epsilon^{abcd}e_b^n\partial_n l_{cd}.
\end{split}
\end{equation}
From here we see that the real and imaginary parts transform independently:
\begin{equation}
\begin{split}
    &\delta^*_{Lor} \text{Re}\, P^a =  l^a_{\;\;b} \text{Re}\, P^b 
    -
    \frac{1}{4}\epsilon^{abcd}e_b^n\partial_n l_{cd},
    \\
     &\delta^*_{Lor} \text{Im}\, P^a =  l^a_{\;\;b} \text{Im}\, P^b 
     +
     \frac{1}{2} e_b^m \partial_m l^{ab}.
\end{split}
\end{equation}
However, these transformation laws contain a non-homogeneous term. These terms resemble the Lorentz transformation law of the spin connection $ \omega_{m|ab}$ \eqref{eq: omega Lor}.
It is therefore natural to redefine these fields as:
\begin{equation}\label{eq: redef}
\begin{split}
    &\text{Re}\, P^a(x) = R^a(x) - \frac{1}{4}\epsilon^{abcd}e_b^m (x)\omega_{m|cd}(x),
    \\
    &\text{Im}\, P^a(x) = D^a(x) + \frac{1}{2}e_b^m (x) \omega_m^{ab}(x).
  \end{split}  
\end{equation}
The newly introduced fields $D^a(x)$ and $R^a(x)$ have homogeneous transformation laws under Lorentz rotations. 
It should be noted that similar redefinitions were also made in the pioneering works of Ogievetsky and Sokatchev on $\mathcal{N}=1$ supergravity \cite{Ogievetsky:1979bb} (see also ref \cite{BK},  page 392).
 
After carrying out this redefinition, it is necessary to reexamine the properties of the fields under Weyl transformations \eqref{eq: Weyl transf}, \eqref{eq: P transf}. From the transformation law of the spin connection under Weyl transformations, we obtain:
\begin{equation}
\delta_d \omega_m^{ab} = -2 e_m^{[a} e^{b]n} \partial_n d
\qquad
\Rightarrow
\qquad
\left[ 
      \begin{gathered} 
        \delta_d \left( \epsilon^{abcd}e_b^m\omega_{m|cd} \right) = -d \left( \epsilon^{abcd}e_b^m\omega_{m|cd} \right), \\ 
        \delta_d \left( e_b^m \omega_m^{ab} \right)
        =
        - d \left( e_b^m \omega_m^{ab} \right)
        +
        3e^{an} \partial_n d. \;\;\;\\ 
      \end{gathered} 
\right.
\end{equation}
The homogeneous terms correspond to those obtained in the eq. \eqref{eq: trans P1}. Their presence results in the following modification of the Weyl transformation law \eqref{eq: P transf} of the field $D_a(x)$:
\begin{equation}\label{eq: D a 0}
    \delta_d D_a(x) = -d(x) D_a(x) - e_a^n(x)\partial_n d(x). 
\end{equation}

\subsection*{Special conformal transformations -- parameter $k^n(x)$} 

At this stage, only one $SU(2)$-singlet transformation parameter remains. It is contained in the superfield $\lambda^{++}$:
\begin{equation}
 \lambda^{++} \Big|_{k_a} = i \theta^{+} \sigma^a \bar{\theta}^{+} k_a(x).
\end{equation}
Since only superfields $\mathcal{H}^{++\hat{\alpha}+}$ and $\mathcal{H}^{(+4)}$ transform under the parameter $\lambda^{++}$, just two of the component fields have nontrivial transformation laws:
\begin{equation}
\begin{split}
    &\delta^*_k P_a(x) \overset{\eqref{eq: redef}}{=} \delta_k R_a(x) + i\delta_k D_a(x) = \frac{i}{2} k_a(x),
    \\
     & \delta^*_k D(x) = e^m_a(x) \partial_m k^a(x) + 2 \text{Im}\, P_a(x) k^a(x) 
     =
     \nabla_a k^a (x) .
   \end{split} 
\end{equation}
Here we use the notation for the covariant derivative introduced in Appendix \ref{eq: spin con}. 

The $k_a$-transformations act on field $D_a$ as an algebraic shift. This allows one to choose the
K-gauge, i.e.~impose~$ D_a(x) = 0$. Thus, the dilatation gauge field completely disappears from the dynamical sector.
The dilatation transformations given by Eq. \eqref{eq: D a 0}
\begin{equation}
    \delta^*_{k, d} D_a(x) = - e^n_a \partial_n d(x) + \dfrac{1}{2} k_a(x)
\end{equation}
lead to an additional condition: to preserve the K-gauge, one must impose a decomposition law
\begin{equation}
k_a(x) = 
2 e_a^m \partial_m d(x) = 2\nabla_a d(x).
\end{equation}
Then the Weyl transformation law for field $D(x)$ takes the form
\begin{equation}
    \begin{split}
    \delta^*_d D(x) &= 
    2 \nabla^a \nabla_a d(x) 
    -
    2d(x) D(x).
    \end{split}
\end{equation}

Comparing the obtained transformation law with the Weyl transformation of the scalar curvature \eqref{eq:scalarR}
\begin{equation}
    \delta_d R(x) = -6 \nabla^a \nabla_a d(x) - 2 d(x) R(x),
\end{equation}
we proceed to redefine field $D(x)$ as
\begin{equation}\label{eq: D red sing}
    D(x) \quad
    \to 
    \quad D(x) - \frac{1}{3} R(x).
\end{equation}
Then the redefined field $D(x)$ has a homogeneous transformation law under Weyl rescaling:
\begin{equation}
    \delta^*_d D(x) = - 2d(x) D(x).
\end{equation}

It should be noted that in the original work \cite{Galperin:1987ek}, the gauge $D(x)=0$ (which implies $\mathcal{H}^{(+4)}=0$) was chosen without imposing the gauge condition $D_a(x)=0$. As explained in the book \cite{18}, such a condition can only be imposed locally. As we will see, the gauge chosen by us is the most convenient one for component analysis.

\subsection*{Final result in $SU(2)$-singlet sector}

\begin{subequations}

The final form of the WZ-type gauge in the $SU(2)$-singlet sector is given by
\begin{equation}\label{eq:WZ}
\boxed{\begin{split}
    &\mathcal{H}^{++m}_{WZ} =  -2i \theta^{+} \sigma^a {\bar{\theta}}^{+} e^m_a, 
\\
    &\mathcal{H}_{WZ}^{++\beta+} = +
    (\theta^+)^2 \bar{\theta}^{+\dot{\alpha}} (\sigma^a)^{\beta}_{\dot{\alpha}} \left(R_a 
    -
    \frac{1}{4}\epsilon^{abcd} \omega_{b|cd}  +
    \frac{i}{2} e_b^m \omega_m^{ab}
    \right)
    +
    (\bar{\theta}^+)^2 \theta^{+\alpha} T_{(\alpha}^{\;\beta)},
\\
    &\mathcal{H}^{++\dot{\beta}+}_{WZ}  = -(\bar{\theta}^+)^2 \theta^{+\alpha} 
    (\sigma^a)^{\dot{\beta}}_{\alpha}
    \left(R_a -  \frac{1}{4}\epsilon^{abcd} \omega_{b|cd}  -
    \frac{i}{2} e_b^m \omega_m^{ab}
    \right)
    +
    (\theta^+)^2 \bar{\theta}^{+\dot{\alpha}} \bar{T}_{(\dot\alpha}^{\;\dot{\beta})},
\\    
    &\mathcal{H}^{(+4)}_{WZ}  = (\theta^+)^4\left( D-\frac{1}{3} R\right).
    \end{split} }
\end{equation}
It is important to emphasize that the correct redefinition of fields has led to the appearance of terms constructed from the spin connection, as well as to the emergence of scalar curvature. These terms play a significant role in the study of the component structure of various $\mathcal{N}=2$ models, in particular the model of the hypermultiplet in the background of the off-shell $\mathcal{N}=2$ conformal supergravity.

Interestingly, similar redefinitions are typical for $\mathcal{N}=2$ higher-spin theories, both the Fronsdal-like \cite{Buchbinder:2021ite} and the conformal ones \cite{Buchbinder:2024pjm}. 
It would be instructive to examine this phenomenon in more detail and to study the connection with the ``gauge formulation'' of higher-spins theories \cite{Vasiliev:1980as}, since the described observations  indicate that this formulation is naturally embedded within the higher-spin analytic prepotentials construction.

Residual gauge freedom is given by the following superfield parameters:
\begin{equation}\label{lam_res_WZ}
\boxed{
\begin{split}
    &\lambda^n_{WZ}  = a^n(x),
\\
    &\lambda^{+\beta}_{WZ}  = \frac{1}{2}\theta^{+\beta} [d(x)+ir(x)] + \theta^{+\alpha} l^{(\beta}_{\;\alpha)}(x), 
\\
    &\bar{\lambda}^{+\dot{\beta}}_{WZ}  = 
    \frac{1}{2}\bar{\theta}^{+\dot{\beta}} [d(x)- ir(x)]
    +
    \bar{\theta}^{+\dot\alpha} \bar{l}^{(\dot{\beta}}_{\;\dot{\alpha})}(x) ,     
\\    
    &\lambda^{++}_{WZ}  = 
 2 i \theta^{+} \sigma^a \bar{\theta}^{+} e_a^m \partial_m d(x). 
\end{split}}
\end{equation}
To summarize, the resulting superparameters have a rather simple form. In particular, they do not depend on the supermultiplet fields (except for the last term in $\lambda^{++}$). In the $SU(2)$ sector, partial gauge fixing will lead to a more complicated dependence of the superparameters on the supermultiplet fields.

For completeness, we provide the full transformation laws of the $SU(2)$-singlet components:
\begin{equation}\label{SU(2)-singlet_transform-s}
\begin{split}
   & \delta^* e_a^n = e^m_a \partial_m a^n - a^n \partial_n e^m_a
    + l_{[a}^{\;\;b]} e^n_b
    - d e_a^{n},
\\
& \delta^* R_a = - a^n \partial_n R_a - \frac{1}{2} e_a^n \partial_n r 
+
l_{[a}^{\;\;b]} R_b 
    -
    d  R_a,
\\
&\delta^* T_{(\alpha\beta)} 
    =
    -
     a^n \partial_n T_{(\alpha\beta)} 
    - \dfrac{1}{2} [d- i r]\,T_{(\alpha\beta)}
    - l^{\;\;\gamma)}_{(\beta}  T_{(\alpha\gamma)} - l_{(\alpha}^{\;\;\gamma)}  T_{(\gamma\beta)},
\\
&\delta^* D = -  a^n \partial_n D  -2dD.
\end{split}
\end{equation}

\end{subequations}

For compactness of formulas, it proves to be convenient subsequently to use the notation
\begin{equation}\label{eq: P field red}
    P_a := R_a 
    -
    \frac{1}{4}\epsilon^{abcd} \omega_{b|cd}  +
    \frac{i}{2} e_b^m \omega_m^{ab}.
\end{equation}



\subsubsection{$SU(2)$ sector}

We now proceed to examine the $SU(2)$ sector, which exhibits the most intricate structure. 
This is due to the fact that $SU(2)$ transformations are realized non-linearly in harmonic superspace.
Our goal is to construct the WZ-preserving gauge transformation parameters by inducing corrections to the initial analytic parameters. According to eq. \eqref{lam_res_1}, the transformation in the $SU(2)$ sector is initially generated by a single harmonic superfield parameter:
\begin{equation}
   \lambda^{++}_{(1)} = \lambda^{(ij)}(x) u^+_i u^+_j.
\end{equation}
The other parameters initially vanish in the $SU(2)$ sector:
\begin{equation}
    \lambda^m_{(1)} = 0,
    \qquad
    \lambda^{+\alpha}_{(1)} = 0,
    \qquad
    \bar{\lambda}^{+\dot{\alpha}}_{(1)} = 0.
\end{equation}
The action of the covariant harmonic derivative $\mathfrak{D}^{++}_{WZ}$ on $\lambda^{++}_{(1)}$ produces terms that violate the Wess-Zumino gauge conditions, see eqs. \eqref{eq: H transf} and \eqref{eq:WZ}. To restore the WZ gauge, the gauge parameters must be augmented by a finite series of compensating corrections:
\begin{equation}
    \lambda^M_{\text{WZ}} = \lambda^M_{(1)} + \lambda^M_{(2)} + \dots + \lambda^M_{(n)}.
\end{equation}
For future convenience, let us isolate the singlet component of the covariant harmonic derivative:
\begin{equation}
    \mathfrak{D}^{++} = 
    \partial^{++}
    +
    \mathfrak{D}^{++}_{singlet}
    +
    (\theta^+)^4 \left\{ V^{n(ij)} u^-_i u^-_j \partial_n + (D-\frac{1}{3}R) \partial^{--} \right\}.
\end{equation}
The singlet part is given by $\mathcal{H}^{++m}_{WZ}$ and $\mathcal{H}^{++\hat{\alpha}+}_{WZ}$, presented in the equation \eqref{eq:WZ}.

This decomposition reveals the operational strategy: the partial harmonic derivative $\partial^{++}$ acts on the higher-order correction $\lambda^M_{(n)}$ to absorb the unwanted terms generated by the background fields acting on $\lambda^M_{(n-1)}$. Crucially, the intermediate operator $\mathfrak{D}^{++}_{\text{singlet}}$ contains no explicit dependence on harmonic variables $u^\pm_i$, which means it preserves the harmonic structure while systematically shifting the $\theta$-power. In particular, its action on the relevant intermediate $\theta$-structures cleanly generates the standard space-time covariant derivative $\nabla_a$:
\begin{equation}
    \mathfrak{D}^{++}_{singlet} \left(  i \theta^+ \sigma^a \bar{\theta}^+ A_a \right)
    = (\theta^+)^4 \left( e^m_a \partial_m + 2 \text{Im} P_a \right) A^a
    =
    (\theta^+)^4  \nabla_a A^a.
\end{equation}

Armed with this iterative machinery and the explicit decomposition of $\mathfrak{D}^{++}$, we are now positioned to systematically evaluate the gauge-preservation conditions across the individual superfield sectors. We begin by applying this setup to the $\mathcal{H}^{(+4)}$ prepotential, where the requirement of maintaining the WZ gauge form will directly determine the first higher-order corrections to the harmonic parameter $\lambda^{++}$ alongside the induced transformations of the component fields.

\medskip

\noindent\underline{\textit{$\mathcal{H}^{(+4)}$  sector}}

\medskip

In this sector, the condition for preserving WZ gauge is given by
\begin{equation}\label{eq: H4 WZ}
    \delta^* \mathcal{H}^{(+4)}_{WZ} = - \lambda^M_{WZ} \partial_M \mathcal{H}^{(+4)}_{WZ} + \mathfrak{D}^{++} \lambda^{++}_{WZ}
    \sim \mathcal{H}^{(+4)}_{WZ}.
\end{equation}
The computation of the covariant harmonic derivative of  $\lambda^{++}_{(1)}$ yields
\begin{equation}
    \mathfrak{D}^{++} \lambda^{++}_{(1)} = 
    \mathfrak{D}^{++}_{singlet} \lambda^{ij} u^+_i u^+_j
    +
    (\theta^+)^4 \Big\{ V^{n(k l)}  \partial_n \lambda^{(ij)} 
    u^-_{k} u^-_{l} u^+_{i} u^+_{j}
    +
    2 (D-\frac{1}{3}R) \lambda^{ij} u^+_i u^-_j\Big\}.
\end{equation}
Let us express the first and third terms as the total partial harmonic derivatives:
\begin{equation}
\begin{split}
    \mathfrak{D}^{++}_{singlet} \lambda^{ij} u^+_i u^+_j
    &=
    \partial^{++} \left(\mathfrak{D}^{++}_{singlet} \lambda^{ij} u^+_i u^-_j\right),
    \\
    2 \left(D-\frac{1}{3}R \right) \lambda^{ij} u^-_i u^+_j &= \partial^{++} 
    \left[ (D-\frac{1}{3}R) \lambda^{ij} u^-_i u^-_j \right].
    \end{split}
\end{equation}
To treat the second term analogously, we decompose the product of harmonics into irreducible components using the following identity:
\begin{equation}
\begin{split}
u^+_{i} u^+_{j} u^-_{k} u^-_{l}
 =\, &
 u^+_{(i} u^+_{j} u^-_{k} u^-_{l)}
 +
 \frac{1}{2} \epsilon_{i (k} u^+_{j} u^-_{l)}
 +
 \frac{1}{3} \big(
 \epsilon_{j k} u^{+}_{(i} u^-_{l)} + \epsilon_{j l} u^{+}_{(i} u^-_{k)}
 \big)
 +
\frac{1}{6} \big(
 \epsilon_{i k} \epsilon_{j l} + 
 \epsilon_{i l} \epsilon_{j k}
 \big)
 . 
 \end{split}
\end{equation}
Applying this decomposition, we obtain
\begin{equation}
\begin{split}
 V^{n(kl)}  \partial_n \lambda^{(ij)} 
    u^+_{i} u^+_{j} u^-_{k} u^-_{l}
    =&\,
     V^{n(kl)}  \partial_n \lambda^{(ij)} 
    u^+_{(i} u^+_{j} u^-_{k} u^-_{l)}
  \\&  +
    \frac{2}{3}
    V^{n \; i)}_{\; (k} \partial_n \lambda^{(kj)} u^+_{i} u^-_{j}
    +
    \frac{1}{3} V^{n(ij)} \partial_n \lambda_{(ij)}.
    \end{split}
\end{equation}
The first two terms can be represented as a total partial harmonic derivative (i.e. as $\partial^{++}(\dots)$), while for the third one nothing else is left but to determine the contribution to the SU(2) transformation of the field $D(x)$:
\begin{equation}\label{eq: D su(2) transf}
    \delta^{*(1)}_{su(2)} D(x) =  \frac{1}{3} V^{n(ij)} \partial_n \lambda_{(ij)}.
\end{equation}
This indicates the need for the field $D(x)$ to be additionally redefined so that it has a trivial SU(2) transformation law.

Collecting the terms representable as total $\partial^{++}$-derivatives, we find that the expression for the $\lambda^{++}_{(2)}$ corrections takes the form
\begin{equation}
\begin{split}
    \lambda^{++}_{(2)} =  \,
   &2 i \theta^{+}\sigma^a\bar{\theta}^{+} e^m_a \partial_m \lambda^{(ij)} u^+_i u^-_j
    -
    (\theta^+)^4 \left(D - \frac{1}{3}R\right) \lambda^{ij} u^-_i u^-_j -
 \\&   -
  \dfrac{1}{3}
    (\theta^+)^4
    \bigg(
  V^{n(ij)} \partial_n \lambda^{(kl)} 
    u^-_{(i} u^-_{j} u^-_{k} u^+_{l)}
    +
    V^{n \; j)}_{\; (k} \partial_n \lambda^{(k l)} u^-_{l} u^-_{j}
    \bigg).
    \end{split}
\end{equation}
Further action of the singlet part of the harmonic derivative on this parameter yields only one new contribution:
\begin{equation}
    \mathfrak{D}^{++}_{singlet} \lambda_{(2)}^{++} 
    = 2(\theta^+)^4 \nabla^a \nabla_a \lambda^{(ij)} u^+_i u^-_j .
\end{equation}
To eliminate this term, we introduce
\begin{equation}
    \lambda^{++}_{(3)} = - (\theta^+)^4 \nabla^a \nabla_a \lambda^{(ij)} u^-_i u^-_j .
\end{equation}
The intermediate expression for the parameter $\lambda^{++}_{WZ}$ is thus given by:
\begin{equation}\label{eq: lambda++ int}
\begin{split}
    \lambda^{++}_{(1)} + \lambda^{++}_{(2)} + \lambda^{++}_{(3)}  =
    & \; \lambda^{(ij)} u^+_i u^+_j
    +
    2 i \theta^{+} \sigma^a \bar{\theta}^{+} e_a^m \partial_m  \lambda^{(ij)} u^+_i u^-_j -
 \\&   -
  \frac{1}{3} 
    (\theta^+)^4
    \bigg(
   V^{n(ij)} \partial_n \lambda^{(kl)} 
    u^-_{(i} u^-_{j} u^-_{k} u^+_{l)}
    +
   V^{n \; j)}_{\; (k} \partial_n \lambda^{(k l)} u^-_{l} u^-_{j}
    \bigg)
    -
 \\&
   - (\theta^+)^4 \left( \nabla^a \nabla_a + D - \dfrac{R}{3} \right) \lambda^{(ij)} u^-_i u^-_j .
    \end{split}
\end{equation}
This parameter preserves the form of WZ gauge, as can be directly checked by substituting it into~\eqref{eq: H4 WZ}.
However, one should bear in mind that the expression for the parameter is not final. This is due to the parameters $\lambda^m, \lambda^{+\hat{\alpha}}$ receiving subsequent corrections that contribute to the transformation law \eqref{eq: H4 WZ}. This will lead, in its turn, to additional contributions to \eqref{eq: lambda++ int}.

\medskip

\noindent\underline{\textit{$\mathcal{H}^{++\hat{\alpha}+}$ sector}}
\medskip
\\We now turn to the condition of Wess—Zumino gauge conservation for $\mathcal{H}^{++\hat{\alpha}+}$:
\begin{equation}
    \delta^*_\lambda \mathcal{H}_{WZ}^{++\hat{\alpha}+} = - \lambda_{WZ}^M \partial_M \mathcal{H}^{++\alpha+}_{WZ} + \mathfrak{D}^{++} \lambda_{WZ}^{+\hat{\alpha}} 
		-
		\lambda_{WZ}^{++} \theta^{+\hat{\alpha}}
        \sim \mathcal{H}_{WZ}^{++\hat{\alpha}+}.
\end{equation}
The $\lambda^{++}$ terms (3.55) produce the contribution
\begin{equation}
   - \theta^{+\alpha} \left(  \lambda^{++}_{(1)} + \lambda^{++}_{(2)} + \lambda^{++}_{(3)}  \right)
    =
  - \theta^{+\alpha} \lambda^{(ij)} u^+_i u^+_j 
    + i (\theta^{+})^2 \bar{\theta}^{+\dot{\beta}} (\sigma^a)^\alpha_{\dot{\beta}} e_a^m \partial_m  \lambda^{(ij)} u^+_i u^-_j.
\end{equation}
which call for a subsequent correction to the  parameter $\lambda^{+\alpha}$:
\begin{equation}
\begin{split}
    \lambda^{+\alpha}_{(2)}
    = &\; \theta^{+\alpha} \lambda^{(ij)}  u^+_i u^-_j 
    - \dfrac{i}{2}
    (\theta^+)^2 \bar{\theta}^{+\dot{\beta}} (\sigma^a)^\alpha_{\dot{\beta}}  e_a^m \partial_m  \lambda^{ij} u^-_i u^-_j.
\end{split}
\end{equation}   
This correction, in turn, generates terms
\begin{subequations}
\begin{equation}
    -  \lambda^{+\hat{\alpha}}_{(2)} \partial^-_{\hat{\alpha}} \mathcal{H}^{++\alpha+}_{WZ}
    =
    - 3 (\lambda^{ij} u^+_i u^-_j)   \mathcal{H}^{++\alpha+}_{WZ},
\end{equation}
\begin{equation}
    \mathfrak{D}^{++}_{singlet} \lambda^{+\alpha}_{(2)}
    =
    i (\theta^+)^2 \bar{\theta}^{+\dot{\beta}} (\sigma^a)_{\dot{\beta}}^{\alpha} e_a^m \partial_m \lambda^{(ij)} u^+_i u^-_j
    +
    (\lambda^{ij} u^+_i u^-_j) \mathcal{H}^{++\alpha+}_{WZ}.
\end{equation}
\end{subequations}
To cancel these terms, we introduce a correction
\begin{equation}
\begin{split}
    \lambda^{+\alpha}_{(3)} 
  =&
    - \dfrac{i}{2} (\theta^+)^2 \bar{\theta}^{+\dot\beta} (\sigma^a)^\alpha_{\dot{\beta}}e^m_a \partial_m \lambda^{ij} u^-_i u^-_j 
 \\&   +
    (\theta^+)^2 \bar{\theta}^{+\dot\beta} P_{\dot{\beta}}^\alpha \lambda^{ij} u^-_i u^-_j
    +
    (\bar{\theta}^+)^2 \theta^{+\beta} T_{(\beta}^{\alpha)} \lambda^{ij} u^-_i u^-_j
    .
\end{split}
\end{equation}
Further action does not lead to any additional corrections. Thus the final form of $\lambda^{+\alpha}$ preserving the WZ gauge form is given by
\begin{equation}\label{eq: lambda+ WZ}
\begin{split}
    \lambda^{+\alpha}_{WZ} 
   = & \,
    \theta^{+\alpha} \lambda^{(ij)}  u^+_i u^-_j 
    - 
    i (\theta^+)^2 \bar{\theta}^{+\dot\beta} (\sigma^a)^\alpha_{\dot{\beta}}e^m_a \partial_m \lambda^{ij} u^-_i u^-_j
\\&    +
    (\theta^+)^2 \bar{\theta}^{+\dot\beta} P_{\dot{\beta}}^\alpha \lambda^{ij} u^-_i u^-_j
    +
    (\bar{\theta}^+)^2 \theta^{+\beta} T_{(\beta}^{\;\alpha)} \lambda^{ij} u^-_i u^-_j.
\end{split}
\end{equation}
The first term in $\lambda^{+\alpha}_{WZ}$ provides extra contribution to $\lambda^{++}_{WZ}$:
\begin{equation}
 \lambda^{++}_{(4)} = 2 (\theta^+)^4 \left(D - \dfrac{1}{3} R\right) \lambda^{(ij)} u^-_i u^-_j.
\end{equation}
Collecting the contributions from \eqref{eq: lambda++ int}, we obtain
\begin{equation}\label{eq: lambda ++ WZ}
    \begin{split}
        \lambda^{++}_{WZ} =& \; \lambda^{(ij)} u^+_i u^+_j
    +
    2 i \theta^{+} \sigma^b \bar{\theta}^{+} e^m_b \partial_m \lambda^{(ij)} u^+_i u^-_j -
 \\&   -
    (\theta^+)^4 
    \bigg(
    \frac{1}{3} V^{n(ij)} \partial_n \lambda^{(kl)} 
    u^-_{(i} u^-_{j} u^-_{k} u^+_{l)}
    +
    \dfrac{1}{2} V^{n \; j)}_{\; (k} \partial_n \lambda^{(k l)} u^-_{l} u^-_{j}
    \bigg)
    -
 \\&
   - (\theta^+)^4 \Big( \nabla^a \nabla_a - D + \dfrac{R}{3} \Big) \lambda^{(ij)} u^-_i u^-_j 
    .
    \end{split}
\end{equation}
In what follows, we show that considering the parameter $\lambda_{WZ}^m$ does not result in any further corrections to expressions \eqref{eq: lambda+ WZ} and \eqref{eq: lambda ++ WZ}.

\smallskip

For completeness, we provide the expression for the tilde-conjugate:
\begin{equation}
    \begin{split}
\bar{\lambda}^{+\dot{\alpha}}_{WZ} = & \,
    \bar{\theta}^{+\dot\alpha} \lambda^{(ij)}  u^+_i u^-_j
    -
    i(\bar{\theta}^+)^2 {\theta}^{+\beta} (\sigma^a)^{\dot\alpha}_{{\beta}}e^m_a \partial_m \lambda^{ij} u^-_i u^-_j
\\&    -
      (\bar{\theta}^+)^2 {\theta}^{+\beta} \bar{P}_{\beta}^{\dot\alpha} \lambda^{ij} u^-_i u^-_j
    +
       ({\theta}^+)^2 \bar\theta^{+\dot\beta} \bar{T}_{(\dot\beta}^{\;\dot\alpha)} \lambda^{ij} u^-_i u^-_j.
    \end{split}
\end{equation}

\noindent\underline{\textit{$\mathcal{H}^{++m}$sector}}

\medskip

The last condition to be analyzed is given by
\begin{equation}
    \delta^*_\lambda \mathcal{H}_{WZ}^{++m} = - \lambda^M_{WZ} \partial_M \mathcal{H}^{++m}_{WZ} +\mathfrak{D}_{WZ}^{++}  \lambda^{m}_{WZ} \label{delta_H^++m} \sim \mathcal{H}_{WZ}^{++m}. 
\end{equation}
Terms to be eliminated are generated as follows:
\begin{equation}
\begin{split}
    - \lambda^{\hat{\alpha}}_{WZ} \partial^{-}_{\hat{\alpha} } \mathcal{H}_{WZ}^{++m}
    =
    & \; 4i \theta^{+}\sigma^a \bar{\theta}^{+} e^m _a\lambda^{(ij)} u_i^+ u^-_j
    -
    4 (\theta^{+})^4 V^{m(kl)} \lambda^{(ij)} u^+_{(i} u^-_{j)} u^-_k u^-_l
    +
    \\
    &
    +
    4 (\theta^+)^4 
    e^m_a \text{Im} P^a {\lambda}^{(ij)}
    u^-_i u^-_j 
    - 
    4 (\theta^+)^4 g^{mn} \partial_n \lambda^{ij} u^+_i u^-_j
    . 
\end{split}
\end{equation}
Here we use the metric tensor $g^{mn} = \eta^{ab}e_a^m e_b^n$.
\\Upon applying the following identity for harmonic monomials:
\begin{equation}
    \begin{split}
        u^+_j u^-_i u^-_k u^-_l = u^-_{(j} u^+_i u^-_k u^-_{l)} + \dfrac{3}{4} \epsilon_{j(i} u^-_k u^-_{l)},
    \end{split}
\end{equation}
it is rewritten as
\begin{equation}\label{eq: var H++m}
    \begin{split}
  - \lambda^{\hat{\alpha}}_{WZ} \partial^{-}_{\hat{\alpha} } \mathcal{H}_{WZ}^{++m}
    =&\,
    -\partial^{++} 
    \Big( -
    2i \theta^{+} \sigma^a \theta^{+} e^m_a \lambda^{(ij)} u^-_i u^-_{j}
    + 
    (\theta^+)^4 \lambda^{(ij)} V^{m(kl)} u^-_{(j} u^-_{i} u^-_k u^-_{l)} 
    \Big)
    \\&
    - 2 (\theta^+)^4 \lambda^{(ij)} V_{i}^{m \; l} u^-_l u^-_{j}
    +
    6 (\theta^+)^4
    e^m_a \text{Im} P^a {\lambda}^{ij}
    u^-_i u^-_j
\\&    - 
    4 (\theta^+)^4 \partial^m \lambda^{(ij)} u^-_i u^-_j
    .
\end{split}
\end{equation}
The last three terms contribute to the SU(2)-transformations of the field $V_m^{ij}$:
\begin{equation}
\begin{split}
    & \delta^{*(1)}_{su(2)} V^{m(ij)} = \; - \big( \lambda^{(kj)} V_{(k}^{m \; i)} + \lambda^{(ki)} V_{(k}^{m \; j)}\big)
    +
    4
    e^m_a \text{Im} P^a {\lambda}^{ij}
    -
    4 \partial^m \lambda^{(ij)}.
\end{split}
\end{equation}
The first line from \eqref{eq: var H++m} gives a correction to $\lambda^m$:
\begin{equation}
\begin{split}
    &\lambda^m_{(2)} = - 2 i \theta^{+} \sigma^a \bar{\theta}^{+} e^m_a \lambda^{(ij)} u^-_i u^-_j 
    +
    (\theta^+)^4 \lambda^{(ij)} V^{m(kl)} u^-_{j} u^-_i u^-_k u^-_{l}. 
\end{split}
\end{equation}
Upon substituting $\lambda^m_{(2)}$ into transformation law \eqref{delta_H^++m}, new terms appear:
\begin{equation}
  \delta^*_{\lambda^m_{(2)}} \mathcal{H}_{WZ}^{++m} 
  =
    -2 (\theta^+)^4 \nabla^a (e_a^m \lambda^{ij}) u^-_i u^-_j
    + 
    2 (\theta^+)^4 e^{na} \partial_n e^m_a \lambda^{(ij)}  u^-_i u^-_j,
    \label{extra}
\end{equation}
which also contributes to the SU(2)-transformations of field $V^{m(ij)}$:
\begin{equation}
\begin{split}
    \delta_{su(2)}^{*(2)} V^{m(ij)} =  - 2  g^{mn} \partial_n  \lambda^{(ij)}
    - 4 e^{m}_{a}  \text{Im} P^a \lambda^{(ij)}
\end{split}
\end{equation}
The resulting SU(2) gauge transformation is given by 
\begin{equation}
\begin{split}
    \delta^*_{su(2)} V^{m(ij)}_{WZ} =&  
    -6\partial^m \lambda^{ij}
    - \big( \lambda^{(kj)} V_{(k}^{m \; i)} + \lambda^{(ki)} V_{(k}^{m \; j)}\big),
    \end{split}
\end{equation}
and the analytic parameter $\lambda^m_{WZ}$ takes the form
\begin{equation}
    \begin{split}
        \lambda^m_{WZ} = \lambda^m_{(2)} =  - 2 i \theta^{+} \sigma^a \bar{\theta}^{+} e^m_a \lambda^{(ij)} u^-_i u^-_j
        +
        (\theta^+)^4 \lambda^{(ij)} V^{m(kl)} u^-_{(j} u^-_i u^-_k u^-_{l)}.
    \end{split}
\end{equation}
One can directly verify that such a form of the parameter preserves Wess–Zumino gauge and does not lead to additional corrections to the $\lambda_{WZ}$-parameters.

The transformation law \eqref{eq: D su(2) transf} of the field $D(x)$ suggests a redefinition
\begin{equation}\label{eq: D SU(2)}
D(x) \quad
\to
\quad
D(x) - \frac{1}{36} V_n^{(ij)} V_{(ij)}^n
\end{equation}
such that the field $D(x)$ becomes an $SU(2)$ invariant scalar.

\subsection*{Final results in $SU(2)$ sector}

\begin{subequations}

As a result, Wess–Zumino gauge in the $SU(2)$ sector takes the following form:
\begin{equation}\label{eq:WZ SU(2)}
\boxed{
\begin{split}
   & \mathcal{H}^{++m}_{WZ} = (\theta^+)^4 V^{m(ij)}(x) u^-_i u^-_j,
   \\
  &  \mathcal{H}^{++\hat{\alpha}+}_{WZ} = 0,
  \\
   & \mathcal{H}^{(+4)}_{WZ} = -\frac{1}{36} (\theta^+)^4\, V_n^{(ij)}(x) V_{(ij)}^n(x).
    \end{split}}
\end{equation}
The analytic parameters in the SU(2) sector that preserve the Wess-Zumino gauge form are given by these expressions:
\begin{equation}\label{eq: lambda SU(2)}
\boxed{
    \begin{split}
        \lambda^m_{WZ} = &- 2 i \theta^{+} \sigma^a\bar{\theta}^{+} e^m_a \lambda^{(ij)} u^-_i u^-_j
        +
        (\theta^+)^4 \lambda^{(ij)} V^{m(kl)} u^-_{(i} u^-_j u^-_k u^-_{l)},
    \\
    \lambda^{+\alpha}_{WZ} =  & \,
    \,
     \,
    \theta^{+\alpha} \lambda^{(ij)}  u^+_i u^-_j 
    - 
    i (\theta^+)^2 \bar{\theta}^{+\dot\beta} (\sigma^a)^\alpha_{\dot{\beta}}e^m_a \partial_m \lambda^{ij} u^-_i u^-_j
   \\& +
    (\theta^+)^2 \bar{\theta}^{+\dot\gamma} P_{\dot{\gamma}}^\alpha \lambda^{ij} u^-_i u^-_j 
    +
    (\bar{\theta}^+)^2 \theta^{+\beta} T_{(\beta}^{\alpha)} \lambda^{ij} u^-_i u^-_j,
    \\
     \lambda^{++}_{WZ} =& \; \lambda^{(ij)} u^+_i u^+_j
    +
    2 i \theta^{+} \sigma^a \bar{\theta}^{+} e^m_a \partial_m \lambda^{(ij)} u^+_i u^-_j -
 \\&   -
    (\theta^+)^4
    \bigg(
    \frac{1}{3} V^{n(ij)} \partial_n \lambda^{(kl)} 
    u^-_{(i} u^-_{j} u^-_{k} u^+_{l)}
    +
    \dfrac{1}{2} V^{n \; j)}_{\; (k} \partial_n \lambda^{(k l)} u^-_{l} u^-_{j}
    \bigg)
    -
 \\&
   - (\theta^+)^4 \Big( \nabla^a \nabla_a - D + \frac{1}{36} V_n^{(ij)} V_{(ij)}^n  + \dfrac{R}{3} \Big) \lambda^{(ij)} u^-_i u^-_j.
    \end{split}}
\end{equation}
Only field $V^{m(ij)}$ has a non-trivial SU(2) transformation law:
\begin{equation}\label{eq: su(2) vector}        
            \delta^*_{su(2)} V^{m(ij)} 
            =  
            -6  \partial^m \lambda^{(ij)}
            +
            \lambda^{i}_{k} V^{m (kj)} + \lambda^{j}_{k} V^{m (ik)}
            .
\end{equation}
\end{subequations}
As shown in Appendix \ref{eq: gauge SU(2)}, these transformations correspond to the standard $SU(2)$ gauge transformations.

 \subsection*{Summary}
 
 As a result, we have imposed the Wess–Zumino-type gauge for $\mathcal{N}=2$ conformal supergravity in the bosonic sector.
 The analytic prepotentials are given by eqs. \eqref{eq:WZ} and \eqref{eq:WZ SU(2)}; the set of analytic gauge superparameters preserving its form is given in \eqref{lam_res_WZ} and \eqref{eq: lambda SU(2)}; the field transformations are specified by eqs. \eqref{SU(2)-singlet_transform-s} and \eqref{eq: su(2) vector}.
 The obtained results are in full agreement with the well-known results in the literature: see, for example, \cite{Freedman:2012zz, Lauria:2020rhc}. 
 
 The most important aspects of our analysis are the structure of field 
$ P_a$ given in eq. \eqref{eq: P field red} and the redefinition of field $D(x)$, see eqs. \eqref{eq: D red sing}, \eqref{eq: D SU(2)}. As we will show below, these redefinitions are necessary for the correct derivation of the component on-shell hypermultiplet action.


	\subsection{Hypermultiplet coupling}

    The hypermultiplet action on the conformal $\mathcal{N}=2$ supergravity background is given by the action in~\eqref{eq: action}.
    After integration by parts with respect to the covariant harmonic derivative, it can be rewritten as
\begin{equation}\label{eq: hyp action}
		S_{hyp} [q, \mathcal{H}] =  - \int d\zeta^{(-4)}\, \tilde{q}^{+} \left( \mathfrak{D}^{++} + \frac{1}{2} \Gamma^{++}\right)q^+,
	\end{equation}	
where we denote a composite analytic superfield $\Gamma^{++}$ as
\begin{equation}
    \Gamma^{++} =  \partial_m \mathcal{H}^{++m}(\zeta) 
    -
    \partial^-_{\hat{\alpha}} 
\mathcal{H}^{++\hat{\alpha}+} (\zeta) 
    +
   \partial^{--} \mathcal{H}^{(+4)}(\zeta). 
\end{equation}
This form will be the most convenient for component analysis. From here one easily obtains the corresponding superfield equations of motion:
	\begin{equation}\label{problem}
		\left(\mathfrak{D}^{++} + \dfrac{1}{2} \Gamma^{++} \right) q^+ = 0. 
	\end{equation}
For illustrative purposes, it is convenient to analyze these equations and the component structure of the action in the singlet and $SU(2)$ sectors separately.

\subsubsection{$SU(2)$ singlet sector}
    
In the $SU(2)$-singlet subsector of the bosonic sector of 4D $\mathcal{N}=2$ conformal supergravity, the Weyl multiplet is described by the set of fields given in~\eqref{eq:WZ}. 
	The differential operator in the EoM then takes the form
	\begin{equation}\label{EOM_diff_oper}
		\begin{split}
			\mathfrak{D}^{++} + \dfrac{1}{2} \Gamma^{++} =& \;
			\partial^{++} 
			- 2i \theta^{+} \sigma^a \bar{\theta}^{+} 
			\hat{\nabla}_a
				+
			(\theta^+)^4 \left( D-\frac{1}{3} R\right) \partial^{--} 
			\\
			&+ \left\{ + (\theta^+)^2 \bar{\theta}^{+\dot{\alpha}} P^{\alpha}_{\dot{\alpha}} 
			+
			(\bar{\theta}^{+})^2\theta^{+\beta} T_{(\beta}^{\alpha)}  \right\} \partial_{\alpha}^- 
			\\&+
			\left\{ -(\bar{\theta}^{+})^2\theta^{+\alpha} \bar{P}^{\dot{\alpha}}_{\alpha}  
			+
			(\theta^+)^2 \bar{\theta}^{+\dot{\beta}} \bar{T}_{(\dot{\beta}}^{\dot{\alpha})} \right\} \bar{\partial}_{\dot{\alpha}}^-.
		\end{split}
	\end{equation}
	Here we have introduced a useful notation of the "hat covariant derivative":
	\begin{equation}\label{eq: hat cov der}
		\hat{\nabla}_a f^i(x) := \left( e_a^m \partial_m  + \frac{1}{2} \partial_m e^m_a -  \text{Im} P_a \right) f^i(x).
	\end{equation}
	The hypermultiplet superfield contains infinitely many component fields in the bosonic sector:
	\begin{equation}\label{eq: hyp exp}
		\begin{split}
			q^+(\zeta_A, u) =&\; F^+ (x, u)  
			+ 
			(\theta^+)^2 M^-(x,u)  
			+
			(\bar{\theta}^+)^2 N^-(x, u) 
		\\&\qquad\qquad	+
			2i \theta^{+} \sigma^a \bar{\theta}^+ A^-_a(x, u)
			+
			(\theta^+)^4 K^{(-3)}(x, u).
		\end{split}
	\end{equation}
The superfield equations \eqref{problem} in WZ gauge  yield the following component equations: 
\begin{equation}\label{eq: conf hyp system}
\begin{cases}
\partial^{++} F^{+}  = 0,
\\
\partial^{++} M^- = 0, 
\\
\partial^{++}  N^- = 0,
\\
\partial^{++}  A^-_a 
			-
			 	\hat{\nabla}_a F^+
			= 0,
\\            
\partial^{++}  K^{(-3)}
			+
			\left( D - \dfrac{1}{3}R \right) \partial^{--} F^+
			+
				\left( \hat{\nabla}_a + 2 \text{Im} P_a \right) A^{-a}
			= 0. 
\end{cases}
\end{equation}
  
The first four equations and a part of the last one are algebraic and thus can be solved explicitly, allowing us to eliminate infinitely many auxiliary fields:
\begin{equation}
    \begin{cases}
F^+(x, u) =  f^i(x) u^+_i,
\\
M^-(x,u) = N^-(x, u)  = 0,
\\
	A^-_a (x, u) =  \hat{\nabla}_a f^i(x) u^-_i, 
\\
    K^{(-3)} (x,u) = 0.
    \end{cases}
\end{equation}

The $(\theta^+)^4$ equation allows one to eliminate the auxiliary field $K^{(-3)} = 0 $
and also leads to a dynamical equation on the complex doublet of scalars:
		\begin{equation}\label{eq: dym singlet}
		\left( D - \dfrac{1}{3} R \right) f^i + 2\bigg( 	\hat{\nabla}_a+  2\text{Im} P^a \bigg) 	\hat{\nabla}_a  f^i = 0.
	\end{equation}
	As a result, after eliminating the auxiliary fields, we obtain an expansion for the hypermultiplet of the form
    \begin{subequations}\label{eq: hyper singlet}
	\begin{equation}
			 q^+(\zeta_A, u)\Big|_{\text{phys}} = \; f^i(x) u^+_i
			+
			2 i \theta^{+} \sigma^a \bar{\theta}^{+} \hat{\nabla}_a
			f^i(x) u^-_i,
		\end{equation}  
as well as the one for its tilde-conjugated counterpart:
        \begin{equation}
			\tilde{q}^+ (\zeta_A, u)\Big|_{\text{phys}} = -\bar{f}^i(x) u^+_i 
			-
			2 i \theta^{+} \sigma^a \bar{\theta}^{+ }
				\hat{\nabla}_a 
			\bar{f}^i (x) u^-_i.
	\end{equation}
\end{subequations}       
As a result of these manipulations, we have solved the algebraic equations for the auxiliary fields, so that the component action of the physical ones in the singlet sector contains the only contribution
\begin{equation}\label{eq: hyper component}
  -  \tilde{q}^{+} \left( \mathfrak{D}^{++} + \frac{1}{2} \Gamma^{++}\right)q^+
    \quad
    \to
    \quad
    - \tilde{q}^{+}\Bigg|_{(\theta^+)^0}  \left[ \left( \mathfrak{D}^{++} + \frac{1}{2} \Gamma^{++}\right)q^+ \right]\Bigg|_{(\theta^+)^4}, 
\end{equation}
leading to the component action of the form
\begin{equation}\label{comp_action_conf_singlet}
    \boxed{ S_{hyp} \Big|_{\text{phys}} =  \int d^4x \left[ \hat{\nabla}^a f^i \hat{\nabla}_a \bar{f}_i - \frac{1}{2} \left(D - \dfrac{1}{3}R\right) f^i \bar{f}_i \right]. }
\end{equation}
By construction, the resulting action is invariant under diffeomorphisms; however, its form differs from the conventional one.
In order to interpret the obtained result, it is necessary to analyze the transformation laws of the hypermultiplet fields.
The hypermultiplet transformations \eqref{eq: gauge transformations} have the infinitesimal form
\begin{equation}\label{eq: hyp infin}
    \delta^*_\lambda q^+ = -\frac{1}{2}  \left( \partial_m \lambda^m_{WZ} - \partial^-_{\hat{\alpha}} \lambda^{+\hat{\alpha}}_{WZ} + \partial^{--}\lambda^{++}_{WZ} \right) q^+
    -
    \lambda^M_{WZ} \partial_M q^+. 
\end{equation}
From this we derive the transformation law for the doublet of scalar fields $f^i(x)$ under translations and Weyl transformations:
\begin{equation}
\begin{split}
    &\delta^* f^{i} = -a^m \partial_m f^i - \dfrac{1}{2} (\partial_m a^m) f^i + d \,f^i,
    \label{f transf law}
\end{split}
\end{equation}
which contains a non-standard weight factor $\partial_m a^m$.
This indicates that the field $f^i$ requires redefinition in order to match the standard transformation law. Comparing with the one for $ e := \det e^a_m$:
\begin{equation}
    \begin{split}
        \delta^* e = - e \, e^a_m \delta^* e^m_a \overset{\eqref{SU(2)-singlet_transform-s}}{=} -e\, \partial_m a^m + e\, e^a_m a^n \partial_n e^m_a + 4 e\, d.
    \end{split}
\end{equation}
we make the following redefinition of the scalar doublet $f^i$ in terms of a new field $\phi^i$\footnote{These redefinitions are analogous to those used in~\cite{Ivanov:2025jdp} for the AdS hypermultiplet.}:
\begin{equation}\label{eq: scalar redef}
    f^i = \sqrt{e} \; \phi^i,
    \qquad
    \delta^*  \phi^i = 
     -a^m \partial_m \phi^i  - d\, \phi^i.
\end{equation}
After this redefinition, we obtain the standard transformation law of the scalar field.

According to the eqs. \eqref{eq: redef} and \eqref{eq: Lor Con},
\begin{equation}
    \text{Im}\, P_a 
    =
    \frac{1}{2} e^m_b \omega_{m|a}^{\;\;\;\;\;\;\;b}
    =
    - \frac{1}{2}
    C_{ab}^b
    =
     - \dfrac{1}{2}e_a^m \left( e_n^b \partial_m e^n_b \right)
    +
    \dfrac{1}{2}\partial_n e^n_a
    =
    \frac{1}{\sqrt{e}}e_a^m \partial_m \sqrt{e}
    +\dfrac{1}{2}
    \partial_n e^n_a. 
\end{equation}
Then for the "hat covariant derivative" we obtain
\begin{equation}\label{f_to_phi_end}
    \begin{split}
        \hat{\nabla}_a f^i =
         \hat{\nabla}_a \left( \sqrt{e} \; \phi^i \right)
        = &\;
        \sqrt{e}\; e^m_a \partial_m \phi^i 
        + 
        \big(
        e^m_a \; \partial_m \sqrt{e} + \dfrac{1}{2} \sqrt{e} \; \partial_m e^m_a - \sqrt{e} \;\text{Im} P_a
        \big) \phi^i 
        \\=& \; \sqrt{e} \; e^m_a \partial_m \phi^i.
    \end{split}
\end{equation}
Substituting this redefinition into formula \eqref{comp_action_conf_singlet}, we obtain
\begin{equation}\label{eq: hyp singlet}
    \begin{split}
        S_{hyp}\Big|^{\text{singlet}}_{\text{phys}} =&\,\int d^4x e \left[  
            g^{mn} \partial_m \phi^i \, \partial_n \bar{\phi}_i
          -\frac{1}{2}
          \left(D- \dfrac{1}{3}R\right)\phi^i \bar{\phi}_i
        \right]. 
    \end{split}
\end{equation}
Thus, the redefinition of the scalar fields has led to the appearance of the integration measure and the inverse metric.


\subsubsection{$SU(2)$ sector}

The contribution to differential operator \eqref{EOM_diff_oper} from the $SU(2)$ sector fields is given by: 
\begin{equation}
    \left(\mathfrak{D}^{++} + \dfrac{1}{2} \Gamma^{++}\right)_{SU(2)} = (\theta^+)^4 \bigg( 
    V^{m(ij)}\partial_m + \dfrac{1}{2} \partial_m V^{m(ij)}
    \bigg)
    u^-_i u^-_j
    - \frac{1}{36} (\theta^+)^4  V^{(ij)}_n V^n_{(ij)} \partial^{--}
    .
\end{equation}
The only modifications to the hypermultiplet system \eqref{eq: conf hyp system} occur 
in the $(\theta^+)^4$ equation. In particular, the dynamical equation of motion \eqref{eq: dym singlet} with SU(2) contributions takes the form: 
\begin{subequations}\label{eq: SU(2) contr}
\begin{equation}
    \begin{split}
       &\left(D - \frac{1}{36} V_n^{(kl)} V_{(kl)}^n  - \dfrac{1}{3}R\right) f^i  
     +
        2\bigg( 	\hat{\nabla}_a+ 2\text{Im} P^a \bigg) 	\hat{\nabla}_a  f^i -
        \\
        &\qquad\qquad\qquad\qquad-
        \dfrac{2}{3}  \left(  V^{m(ij)} \partial_m    
        +
        \frac{1}{2}
         \partial_m  V^{m(ij)}   
        \right) f_j
          = 0
    \end{split}
\end{equation}
while the relation for $K^{(-3)}$, which is now rendered nontrivial, becomes as follows:
\begin{equation}
    K^{(-3)} (x,u) =  - \dfrac{1}{3}
    \left(  V^{m (ij)} \partial_m f^k 
    + \dfrac{1}{2} \partial_m V^{m(ij)} f^k \right) \; u^-_{(i} u^-_j u^-_{k)}.
\end{equation}
\end{subequations}
Then the on-shell hypermultiplet expansion \eqref{eq: hyper singlet} acquires an extra nonzero term $K^{(-3)}$:
\begin{equation}
\begin{split}
			 q^+(\zeta_A, u)\Big|_{\text{phys}} = \; &f^i(x) u^+_i
			+
			2 i \theta^{+} \sigma^a \bar{\theta}^{+} \hat{\nabla}_a
			f^i(x) u^-_i
        \\&    - \dfrac{1}{3} (\theta^+)^4
    \left(  V^{m (ij)} \partial_m f^k 
    + \dfrac{1}{2} \partial_m V^{m(ij)} f^k\right) \; u^-_{(i} u^-_j u^-_{k)}.
    \end{split}
	\end{equation}
As in the singlet sector case, inserting it into the action contributes only one term (see eq. (3.87)), and after substituting $f^i = \sqrt{e} \phi^i$ (see eq. \eqref{eq: scalar redef}) we finally arrive at the following SU(2) contribution:
\begin{equation}\label{eq: SU(2) contr} 
\begin{split}
        S_{hyp}\Big|_{SU(2)} =
        \int d^4x\, e\;\Big( 
        -
        \frac{1}{6} V^{m(ij)} \big( \phi_i \partial_m \bar{\phi}_j - \bar{\phi}_j \partial_m \phi_i\big)  - \dfrac{1}{72} V^n_{(ij)} V_n^{(ij)} \phi^k \bar{\phi}_k\Big).
    \end{split}
\end{equation}    
    
The $SU(2)$ transformation law for the doublet of scalar fields can be derived from \eqref{eq: hyp infin} using local $SU(2)$ parameters \eqref{eq: lambda SU(2)}. 
From this, we find the transformation law for the theta-independent part of the hypermultiplet superfield:
\begin{equation}
    \delta^*_{su(2)} F^+ 
    =
    + \left( \lambda^{(ij)} u^+_i u^-_j \right) F^+
    -
    \left( \lambda^{(ij)} u^+_i u^+_j  \right) \partial^{--} F^+.
\end{equation}
Then we obtain that the scalar doublet has a homogeneous transformation law:
\begin{equation}
\delta^*_{su(2)} f^i(x)
=  \lambda^{(i}_{\;\;j)}(x) f^j (x)
\quad
\Leftrightarrow
\quad
\delta^*_{su(2)} \phi^i(x)
=  \lambda^{(i}_{\;\;j)}(x) \phi^j (x).
\end{equation}
Taking into account the gauge transformation law of the $SU(2)$ field \eqref{eq: su(2) vector}, it is natural to introduce an $SU(2)$ covariant derivative:
\begin{equation}
    \mathcal{D}_m \phi^i := \partial_m \phi^i 
    +
    \frac{1}{6} (V_m)^{(i}_{\;\;j)} \phi^j;
\qquad
    \delta^*_{su(2)}  \mathcal{D}_m \phi^i 
    =
    \lambda^{(i}_{\;\;j)} \mathcal{D}_m \phi^j. 
\end{equation}
Then summing (3.95) and (3.99) gives the full on-shell hypermultiplet action:
\begin{equation}\label{eq: hyp conformal}
  \boxed{
        S_{hyp}\Big|_{\text{phys}} =\,\int dx^4 e \left[  
            g^{mn} \mathcal{D}_m \phi^i \, \mathcal{D}_n \bar{\phi}_i
          -\frac{1}{2}
          \left(D- \dfrac{1}{3}R\right)\phi^i \bar{\phi}_i
        \right]. }
\end{equation}

\subsection*{Summary}

As a result, we have obtained a non-minimal Weyl-invariant coupling of the scalar doublet to gravity.
We note once again that the appearance of the non-minimal term $-\frac{1}{6}R\phi^i \bar{\phi}_i$ is a direct corollary of the sequential and careful  redefinition of the fields in Wess–Zumino gauge. 

\smallskip

Let us note that the equation of motion for the auxiliary field $D(x)$ takes the following form: 
\begin{equation}
	\phi^i \bar{\phi}_i = 0.
\end{equation}
This means that we cannot consistently fix a non-trivial vacuum expectation value for the scalar field, which is necessary to obtain the Einstein action via conformal compensators.  This precisely explains why two compensators are needed to construct $\mathcal{N}=2$ Einstein supergravity.

\subsection{Vector compensator contribution}

Let us now introduce a second conformal compensator in a form of some vector superfield denoted as $\mathcal{H}^{++5}$ by considering an extra central-charge coordinate $x^5$ and adding a term $\mathcal{H}^{++5} \partial_5$ to the covariant harmonic derivative \eqref{eq: cov der}. This way, one obtains the vector prepotential $\mathcal{H}^{++5}$ on the $\mathcal{N}=2$ Weyl supergravity background. In this subsection, we describe the Wess–Zumino-type gauge for its vector subsector and obtain the latter's contribution to the on-shell component action \eqref{eq: hyp conformal}, which adds the interaction of the hypermultiplet with the background vector superfield\footnote{The notation is chosen to directly pair with the one in the Section \ref{sec: Ein sg}. As in the Einstein supergravity case, the so-introduced coordinate $x^5$ can be related to the central charge.}.

\smallskip

We start with the infinitesimal transformations for the $\mathcal{N}=2$ vector multiplet prepotential $\mathcal{H}^{++5}$:
\begin{equation}\label{eq: H++5 transf}
	\delta^*_\lambda \mathcal{H}^{++5} := \mathcal{H}^{\prime++5}(\zeta) - \mathcal{H}^{++5}(\zeta)  = \mathfrak{D}^{++} \lambda^5 - \lambda^M_{WZ} \partial_M \mathcal{H}^{++5}.
\end{equation}	
 In flat harmonic superspace, zeroing out the pure gauge degrees of freedom gives the following form for Wess-Zumino gauge:
\begin{equation}
	\begin{split}
		\mathcal{H}^{++5}_{WZ}(\zeta) \Big|_{flat} = \;&i (\theta^+)^2 C(x) - i (\bar{\theta}^+)^2 \bar{C}(x)
		-
		2i \theta^{+}\sigma^a\bar{\theta}^{+} B_{a}(x)
		+
		(\theta^{+})^4 S^{(ij)}(x) u^-_i u^-_j,
	\end{split}
\end{equation}	
with the residual gauge freedom parametrized by the only parameter
\begin{equation}
\label{eq: b transf}
	\lambda^5_{(0)} = b(x).
\end{equation}	
However, in the presence of the $\mathcal{N}=2$ supergravity, further investigation is required. Indeed, according to eq. \eqref{eq: H++5 transf}, the covariant harmonic derivative and the translational terms in the $SU(2)$ sector generate additional contributions
\begin{subequations}
	\begin{equation}
		\mathfrak{D}^{++}_{WZ} 	\lambda^5_{(0)} 
		=
		-2i \theta^+ \sigma^a \bar{\theta}^+ e_a^m \partial_m b
		+
		(\theta^+)^4 V^{m(ij)}  \partial_m b\,  u^-_i u^-_j,
	\end{equation}	
	\begin{equation}
		\begin{split}
			- \lambda^{M}_{WZ} \partial_M 
			\mathcal{H}^{++5} 
			=&
			-
			\partial^{++} \left( 	\mathcal{H}^{++5}   \lambda^{--}  \right)
			+
			2 (\theta^+)^4 S^{(ik}  \lambda_k^{j)} u^-_i u^-_j
			\\
			& -  \left(  \mathfrak{D}^{++}_{WZ}  \mathcal{H}^{++5}  \right)  \lambda^{--} 
			-
			4 (\theta^+)^4 B^a e_a^m \partial_m  \lambda^{--}, 
		\end{split}
	\end{equation}	
\end{subequations}
where we denote $\lambda^{--} :=  \lambda^{ij} u^-_iu^-_j $.

For them to be cancelled, we introduce a correction to the analytic superparameter $\lambda^5$
\begin{equation}
	\lambda^5_{(1)} = \mathcal{H}^{++5}  \lambda^{--}, 
\end{equation}	
which generates new terms
\begin{equation}
	\mathfrak{D}^{++}_{WZ}	\lambda^5_{(1)} 
	=
	\partial^{++} \left( 	\mathcal{H}^{++5}  \lambda^{--}  \right)
	+
	\left(\mathfrak{D}^{++}_{WZ}	 \mathcal{H}^{++5}  \right) \lambda^{--} 
	- 2  (\theta^+)^4 B^a e_a^m \partial_m \lambda^{--}.	
\end{equation}	
Also, we deduce the transformation laws for some of the component fields:
\begin{equation}
	\begin{split}
		&\delta_b B_a = e_a^m \partial_m b,
		\\
		&\delta_{su(2)} D^{(ij)}
		=
		2 S^{(ik} \lambda_k^{j)}
		- 6 B^a e_a^m \partial_m \lambda^{(ij)}.	
	\end{split}
\end{equation}	
It becomes clear that to obtain the triplet of auxiliary scalar fields, one needs to perform a redefinition
\begin{equation}\label{eq: D redef}
	S^{(ij)}(x) \quad\to \quad
	S^{(ij)} (x)
	+
	B^a(x)  V_a^{(ij)}(x).
\end{equation}	
So, the final gauge-preserving form of the parameter $\lambda^5_{WZ}$ is given by
\begin{equation}\label{eq: res WZ vector}
	\lambda^5_{WZ} =  	\lambda^5_{(0)} +  	\lambda^5_{(1)} =  b
	+
	\mathcal{H}_{WZ}^{++5}  \lambda^{--}, 
\end{equation}	
where
\begin{equation}
\begin{split}
   \mathcal{H}_{WZ}^{++5}(\zeta) = &\;i (\theta^+)^2 C(x) - i (\bar{\theta}^+)^2 \bar{C}(x)
		-
		2i \theta^{+}\sigma^a\bar{\theta}^{+} B_{a}(x)
		+
		\\&
        +(\theta^{+})^4\left[ S^{(ij)} (x) +
	    B^a(x)  V_a^{(ij)}(x)\right] u^-_i u^-_j.
\end{split}
\end{equation}

In summary, the transition from flat harmonic superspace to a curved $\mathcal{N}=2$ Weyl supergravity background non-trivially deforms the Wess–Zumino gauge structure. The mixing between the vector prepotential $\mathcal{H}^{++5}_{WZ}$ and the background $SU(2)$ gauge field requires a correction $\lambda^5_{(1)}$ to the analytic gauge parameter. 
Crucially, the shift \eqref{eq: D redef} isolates the true, gauge-covariant triplet of auxiliary scalar fields $D^{(ij)}(x)$. By disentangling the vector multiplet components from the pure supergravity background fields, this formulation guarantees that the subsequent construction of the invariant action, as well as its coupling to matter multiplets, can be performed in a clean, structurally transparent manner.

Under the residual gauge freedom in the $SU(2)$-singlet sector \eqref{lam_res_WZ} and \eqref{eq: res WZ vector} the component fields of the vector multiplet transform as
\begin{equation}\label{eq: vector sm transf}
		\begin{split}
			&\delta^* C = - a^n\partial_n C  - (d+ir) C,
			\\
			&\delta^* B_a =  e_a^m \partial_m b  - a^n\partial_n B_a
			+ l_{[a}^{\;\;b]} B_b
			- d B_a,
			\\
			&\delta^* S^{(ij)}  = - a^n\partial_n  S^{(ij)}  - 2 d S^{(ij)} 
			+
			2 \lambda^{(i}_kS^{j)k}.
		\end{split}	
	\end{equation}	
Since the complex scalar field $C(x)$ transforms with a shift under the dilatation parameter $d$ and the chiral $U(1)_R$ parameter $r$, it can be set to a nonzero constant in the $SU(2)$-singlet sector, thereby verifying its physical role as a conformal compensator.

\smallskip

We now turn to the study of the component action of the hypermultiplet in the presence of a vector multiplet.
The operator in the hypermultiplet equations \eqref{problem} takes the form
\begin{equation}
	\begin{split}
		\mathfrak{D}^{++} + \dfrac{1}{2} \Gamma^{++} =& \;
		\partial^{++} 
		-
		2i \theta^{+} \sigma^a\bar{\theta}^{+} 
		\hat{\nabla}_a
		\\
        &+ 
        i \left[ (\theta^+)^2 C
		-
		(\bar{\theta}^+)^2
		\bar{C} \right] \partial_5 
		+
		\\&+
		\left\{+ (\theta^+)^2 \bar{\theta}^{+\dot{\alpha}} P^{\alpha}_{\dot{\alpha}} 
		+(\bar{\theta}^{+})^2\theta^{+\beta}
		T_{(\beta}^{\alpha)} 
		\right\} \partial_{\alpha}^- 
		\\&+
		\left\{ - (\bar{\theta}^{+})^2\theta^{+\alpha} \bar{P}^{\dot{\alpha}}_{\alpha}  
		+
		(\theta^+)^2 \bar{\theta}^{+\dot{\beta}}             \bar{T}_{(\dot{\beta}}^{\dot{\alpha})}
		\right\} \bar{\partial}_{\dot{\alpha}}^-
		\\&+
		(\theta^+)^4 
		\left\{
		\left(S^{(ij)}
		+ 
		V^{m(ij)} e_m^a B_a\right) \partial_5 
		+
		V^{m(ij)} \partial_m
		+ 
		\dfrac{1}{2} \partial_m V^{m(ij)}
		\right\}
		u^-_i u^-_j
		\\&+
		(\theta^+)^4  \left( D - \frac{1}{3}R	-
		\frac{1}{36} V_n^{(ij)} V^n_{(ij)}\right) \partial^{--},
	\end{split}
\end{equation}	
where we have introduced a "hat covariant derivative" with $U(1)$ connection, defined as
\begin{equation}
	\hat{\nabla}_a 
    :=
    e_a^m \partial_m 
	+
	\frac{1}{2} \partial_m e^m_a - \text{Im} P_a +  B_a \partial_5.
\end{equation}	
It differs from the "hat covariant derivative" \eqref{eq: hat cov der} by the $U(1)$ graviphoton contribution.

To couple to the vector multiplet, we introduce an additional exponential $x^5$-dependence to the hypermultiplet superfield:
\begin{equation}\label{eq: hyper x5}
    \mathbf{q}^+ (\zeta, u, x^5) = e^{im_qx^5} q^+(\zeta, u),
\end{equation}
where the real parameter $m_q$ corresponds to the hypermultiplet mass.

\smallskip

Then the component hypermultiplet EoM \eqref{eq: conf hyp system} itself is rewritten as a system
\begin{equation}\label{EoM_system_vector}
	\begin{cases}
		\partial^{++} F^{+}  = \; 0,
		\\
		\partial^{++} M^- -  m_q C\,   F^{+} = 0, 
		\\
		\partial^{++}  N^- + m_q \bar{C}\,   F^{+} = 0 ,
		\\
		\partial^{++}  A^-_a  -  \hat{\nabla}_a F^+
		= 0,
		\\            
		\partial^{++}  K^{(-3)} + 2\left( \hat{\nabla}_a + 2 \text{Im} P_a \right) A^{-a} \;
		+ m_q \left(  \bar{C} M^-  - C N^- \right)
		\\
		\qquad +
		\left\{
		im_q \left(S^{(ij)}
		+ 
		V^{m(ij)} e_m^a B_a\right) 
		+
		V^{m(ij)}\partial_m
		+
		\dfrac{1}{2} \partial_m V^{m(ij)}
		\right\}
		u^-_i u^-_j F^+ = 0
		,
	\end{cases}
\end{equation}
The algebraic part of equations \eqref{EoM_system_vector} can be solved explicitly:
\begin{equation}
	\begin{cases}
		F^+(x, u) =  f^i(x) u^+_i,
		\\
		M^-(x,u) =  + m_q C f^i(x) u^-_i ,
		\\
		N^-(x, u)  =- m_q \bar{C} f^i(x) u^-_i ,
		\\
		A^-_a(x,u) =  \hat{\nabla}_a f^i(x) u^-_i. 
	\end{cases}
\end{equation}
The last equation in the system \eqref{EoM_system_vector} splits into a non-trivial algebraic equation on $K^{(-3)}$:
\begin{equation}
	K^{(-3)}  = - \dfrac{1}{3}\left\{
	im_q  (S^{(ij)}
	+ 
	V^{m(ij)} e_m^a B_a)
	+
	V^{m(ij)}\partial_m
	+
	\dfrac{1}{2} \partial_m V^{m(ij)}
	\right\} f^k u^-_{(i} u^-_{j} u^-_{k)},
\end{equation}
and a dynamical equation on the complex doublet of scalars:
\begin{equation}
	\begin{split}
		&
		2 \left( \hat{\nabla}_a + 2 \text{Im} P_a \right) \check{\nabla}^a f^i
		+
		2 m^2_q |C|^2  f^i 
		+
		\\&
		-
		\dfrac{2}{3}
		\left\{
		im_q S^{(ij)}
		+
		V^{m(ij)}\left( \partial_m + im_q e^a_m B_a \right) 
		+
		\dfrac{1}{2} \partial_m V^{m(ij)}
		\right\}
		f_j 
		= 0.
	\end{split}
\end{equation}

As a result, after eliminating the auxiliary fields, we obtain the expression for the hypermultiplet in the physical sector:
\begin{equation}
	\begin{split}
		q^+(\zeta_A, u)\Big|_{\text{phys}} =& \; f^i u^+_i
		+
		m_q (\theta^+)^2  C  f^i u^-_i
		-
		m_q
		(\bar{\theta}^+)^2 \bar{C} f^i u^-_i
		+
		2 i \theta^{+} \sigma^a\bar{\theta}^{+} \hat{\nabla}_a f^i u^-_i
		\\&
		- \dfrac{1}{3}
		(\theta^+)^4
		\left\{
		im_q  (S^{(ij)}
		+ 
		V^{m(ij)} B_m)
		+
		V^{m(ij)}\partial_m
		+
		\dfrac{1}{2} \partial_m V^{m(ij)}
		\right\} f^k u^-_{(i} u^-_{j} u^-_{k)} .
	\end{split}
\end{equation}
Then the contribution to the component hypermultiplet action in the physical sector takes the form
\begin{equation}
	\begin{split}
		S_{hyp}\Bigg|_{\text{phys}} 
		=
		\int d^4x \, & \Bigg[ \hat{\nabla}^a f^i \hat{\nabla}_a \bar{f}_i
		-
		m_q^2 |C|^2 f^i \bar{f}_i
		+
		\frac{im_q}{3} S^{(ij)} f_i \bar{f}_j
		-
		\frac{1}{2}\left( D -\frac{1}{3}R \right) f^i \bar{f}_i
		\\&
        -
		\frac{1}{6} V^{m(ij)} \left( \partial_m \bar{f}_i f_j
		-
		\bar{f}_i \partial_m f_j
		\right)
		+
		\frac{im_q}{3} V^{m(ij)} B_m \bar{f}_i f_j
		-
		\frac{1}{72} V^{n}_{(ij)} V_n^{(ij)} f^k \bar{f}_k
		\Bigg].
	\end{split}
\end{equation}	 
After the scalar field redefinition
\begin{equation}
	f^i \quad 
	\to
	\quad
	\sqrt{e} \phi^i
\end{equation}	
we finally arrive at
\begin{equation}\label{conf final action: vector}
    S_{hyp}\Big|_{\text{phys}} 
	=
	\int d^4x e \left[ g^{mn} \mathcal{D}_m \phi^i \mathcal{D}_n \bar{\phi}_i 
	-
	m_q^2 |C|^2 \phi^i \bar{\phi}_i
	+
	\frac{im_q}{3} S^{(ij)} \phi_i \bar{\phi}_j
	-
	\frac{1}{2}\left( D -\frac{1}{3}R \right) \phi^i \bar{\phi}_i
	\right].
\end{equation}	
Here we have introduced an $SU(2)$ covariant derivative
\begin{equation}
	\mathcal{D}_m \phi^i 
	=
	\left( \partial_m + im_q B_m  \right) \phi^i 
	+
	\frac{1}{6} (V_m)^i_{\;j} \phi^j ,
\end{equation}	
which covariantly transforms under the $U(1)$ and $SU(2)$ transformations, see eq. \eqref{eq: b transf} and eq.
\eqref{eq: su(2) vector}.
The differences between the obtained action and the action \eqref{eq: hyp conformal} is the presence of the graviphoton contribution in the covariant derivative and in the interaction of the physical and the auxiliary scalars with the hypermultiplet ones.


\section{$\mathcal{N}=2$ Einstein supergravity multiplet}\label{sec: Ein sg}

We now turn to the Einstein supergravity multiplet.
Similliarly to the conformal case, the $\mathcal{N}=2$ Einstein supergravity is described by a covariant harmonic derivative
\begin{equation}
    \mathbb{D}^{++} = \partial^{++} + H^{++m} \partial_m 
    +
    H^{++\hat{\alpha}+} \partial^-_{\hat{\alpha}}
    +
     H^{++\hat{\alpha}-} \partial^+_{\hat{\alpha}}
    +
      H^{++5} \partial_5,
\end{equation}
where the supervielbeins have transformation laws of the form (compare with \eqref{eq: H transf}):
\begin{equation}\label{eq: H E transf}
	\begin{split}
		\delta_\lambda H^{++m} &= \mathbb{D}^{++}  \lambda^{m},
		\\ 
		\delta_\lambda H^{++\hat{\alpha}+} &= \mathbb{D}^{++} \lambda^{+\hat{\alpha}},
		\\
		\delta_\lambda H^{++\hat{\alpha}-} &= \mathbb{D}^{++} \lambda^{-\hat{\alpha}},
		\\
		\delta_\lambda 
		H^{++5} &= \mathbb{D}^{++} \lambda^{5}.
	\end{split}	
\end{equation}	
As we have already discussed in Section \ref{eq: sec Einstein sg gg}, these transformations can be derived from \eqref{eq: H transf} by setting the parameter $\lambda^{++}$
 to zero.

\subsection{Wess-Zumino-type gauge}

To fix Wess-Zumino gauge, we follow the same strategy as in the conformal case. For the reader's convenience, we will present some intermediate steps that illustrate important differences.

\smallskip

\textbf{1.} In the zeroth order approximation, the supervielbein variations are
\begin{equation}
\begin{split}
    \delta^{(0)} H^{++m} &= \partial^{++} \lambda^m,
   \\
    \delta^{(0)} H^{++\alpha+} &= \partial^{++} \lambda^{+\alpha},
   \\
     \delta^{(0)} H^{++\dot{\alpha}+} &= \partial^{++} \bar{\lambda}^{+\dot{\alpha}},
\\
     \delta^{(0)} H^{++5} &= \partial^{++} \lambda^{5}.
\end{split}
\end{equation}
Analogously to the previous section, we expand the RHS and LHS and then match the degrees of freedom. 
The following components of the superparameters are zeroed out when subject to the action of $\partial^{++}$:
\begin{equation}\label{lam_res}
\begin{split}
    &\lambda^n_{(0)} = a^n(x),
\\
    &\lambda^{+\beta}_{(0)} = \theta^{+\beta} l(x)
    +
    \theta^{+\alpha} l^{(\beta}_{\;\alpha)}(x) 
    + \bar{\theta}^{+\dot{\alpha}} n^\beta_{\dot\alpha}(x) , 
\\
    &\bar{\lambda}_{(0)}^{+\dot{\beta}} = 
    \bar{\theta}^{+\dot{\beta}} \bar{l}(x)
    +
    \bar{\theta}^{+\dot\alpha} \bar{l}^{(\dot{\beta}}_{\;\dot{\alpha})}(x)
    +
    \theta^{+{\alpha}} n_\alpha^{\dot\beta}(x) ,     
\\    
    &\lambda^{5}_{(0)} = b(x). 
\end{split}
\end{equation}
The others can be used to zero out most of the components of the prepotentials, leaving 
\begin{equation}
\begin{split}
    &H^{++n}_{(0)} = (\theta^+)^2 C^n(x)  +(\bar{\theta}^+)^2 \bar{C}^n(x) - 2i \theta^{+} \sigma^a\bar{\theta}^{+} e^n_a(x) + 
    (\theta^+)^4 V^{n(ij)}(x) u^-_i u^-_j, 
\\
    &H^{++\beta+}_{(0)} = +
    (\theta^+)^2 \bar{\theta}^{+\dot{\alpha}} P^{\beta}_{\dot{\alpha}}(x) 
     +
    (\bar{\theta}^+)^2 \theta^{+\beta} T(x) 
    +
    (\bar{\theta}^+)^2 \theta^{+\alpha} T_{(\alpha}^{\;\beta)}(x),
\\
    &H^{++\dot{\beta}+}_{(0)} = -(\bar{\theta}^+)^2 \theta^{+\alpha} 
    \bar{P}^{\dot{\beta}}_{\alpha}(x) 
    +
    (\theta^+)^2 \bar{\theta}^{+\dot{\beta}} \bar{T}(x)
    +
    (\theta^+)^2 \bar{\theta}^{+\dot{\alpha}} \bar{T}_{(\dot\alpha}^{\;\dot{\beta})}(x),
\\    
    &H^{++5}_{(0)} = (\theta^+)^2 C(x) 
    +
    (\bar{\theta}^+)^2 \bar{C}(x)
    -
    2i \theta^{+}\sigma^a \bar{\theta}^{+} B_a (x)
    +
    (\theta^+)^4 S^{(ij)}(x)u^-_i u^-_j.
    \end{split}
\end{equation}
Note that superfields $H^{++n}, H^{++5}, \lambda^n, \lambda^{5}$ are real with respect to harmonic tilde-conjugation. This leads to the corresponding reality conditions for their components.
\bigskip

\noindent\textbf{2.} As in the conformal case, using parameter $n^a$, we obtain the transformation law
\begin{equation}
    \delta_n C^m (x) = 
    -
    2i \bar{n}^a (x) e_a^m(x)
\end{equation}
and so we can fix $C^m(x) = \bar{C}^m(x) = 0$.

\smallskip

Moreover, considering the full transformation law of field $C(x)$, written as
\begin{equation}
\begin{split}
        \delta^* C (x) &= - 2 l(x) C(x) - a^m(x) \partial_m C(x),
 \end{split}   
\end{equation}
allows us to set $C(x) = i$ using the parameters $d(x) = \frac{1}{2}\text{Re}\, l(x)$ and $r(x) = \frac{1}{2}\text{Im}\, l(x)$, which corresponds to flat $\mathcal{N}=2$ superspace with a nontrivial cental charge \cite{18}. This establishes the superfield $H^{++5}$ as a conformal compensator, since it  compensates the Weyl and R-symmetry transformations\footnote{In the original work \cite{Galperin:1987em}, another gauge $C(x) = e(x)$ is fixed. Our sequential gauge-fixing procedure shows that their gauge choice is incompatible with the field transformations.}.

\medskip

\noindent\textbf{3.}  Thus we obtain Wess-Zumino gauge (with no full physical redefinitions performed yet)
\begin{equation}\label{eq: Eins WZ}
	\boxed{
\begin{split}
    &H^{++n}_{WZ} =  -2i \theta^{+} \sigma^a \bar{\theta}^{+} e^n_a(x) + 
    (\theta^+)^4  V^{n\,ij}(x) u^-_i u^-_j, 
\\
    &H^{++\beta+}_{WZ} =
    +
    (\theta^+)^2 \bar{\theta}^{+\dot{\alpha}} P^{\beta}_{\dot{\alpha}}(x)  
    +
    (\bar{\theta}^+)^2 \theta^{+\beta} T(x) 
    +
    (\bar{\theta}^+)^2 \theta^{+\alpha} T_{(\alpha}^{\;\beta)}(x),
\\
    &H^{++\dot{\beta}+}_{WZ} = - (\bar{\theta}^+)^2 \theta^{+\alpha} 
    \bar{P}^{\dot{\beta}}_{\alpha}(x) 
    +
    (\theta^+)^2 \bar{\theta}^{+\dot{\alpha}} \bar{T}(x) 
    +
    (\theta^+)^2 \bar{\theta}^{+\dot{\alpha}} \bar{T}_{(\dot\alpha}^{\;\dot{\beta})}(x),
\\    
    &H^{++5}_{WZ} = i \left[ (\theta^+)^2 -
    (\bar{\theta}^+)^2 \right]
    -
    2i \theta^{+} \sigma^a \bar{\theta}^{+} B_a(x) 
    +
    (\theta^+)^4 S^{(ij)}(x) u^-_i u^-_j
    \end{split} }
\end{equation}
and a finite set of residual gauge parameters:
\begin{equation}\label{lam_res_WZ Ein}
\boxed{ \begin{split}
    &\lambda^n_{WZ} = a^n(x),
\\
    &\lambda^{+\beta}_{WZ} =
    l^{(\beta}_{\;\;\alpha)}(x)\theta^{+\alpha} , 
\\
    &\bar{\lambda}_{WZ}^{+\dot{\beta}} = 
\bar{l}^{(\dot{\beta}}_{\;\;\dot{\alpha})}(x) \bar{\theta}^{+\dot\alpha}  ,     
\\    
    &\lambda^{5}_{WZ} = b(x). 
\end{split}}
\end{equation}
We redefine field $P_a(x)$ as in the conformal case, see eq. \eqref{eq: redef}.

The field $S^{ij}$ transforms under the gauge transformations as 
\begin{equation}
    \delta^* S^{(ij)} = - a^n \partial_n \tilde{S}^{(ij)} + V^{m(ij)} \partial_m b,
\end{equation}
which hints on the shift
\begin{equation}
    S^{(ij)} \quad \to
    \quad
    S^{(ij)}
    + 
    V^{m(ij)} e_m^a B_a
\end{equation}
so that the redefined field $S^{(ij)}$ 
has a trivial $U(1)$ transformation law
\begin{equation}
    \delta^* \left(  S^{(ij)}
    + 
    V^{m(ij)} e_m^a B_a \right)
    =
   -a^n \partial_n  \left(  S^{(ij)}
    + 
    V^{m(ij)} e_m^a B_a \right)
    +
     V^{m(ij)} \partial_m b.
\end{equation}
and thus indeed corresponds to an auxiliary field.

\subsection*{Final result}

The final form of the WZ-type gauge in the bosonic sector is given by \eqref{eq: Eins WZ}, where field $P^a$ is given by
\begin{equation}
    P^a (x) := R^a(x) - \frac{1}{4}\epsilon^{abcd}\omega_{b|cd}(x)  - iD^a(x) + \frac{i}{2}e_b^m (x)\omega_m^{ab} (x).
\end{equation}
Several differences from the conformal setup stand out: due to the lack of the $R$ symmetry transformation parameter, which is compensated by the scalar from the vector multiplet, $R_a$
ceases to be the Weyl gauge field. Moreover, as we cannot fix the K-gauge $D_a=0$ for $D_a(x)$, it remains in the final expression for $P_a$
as an auxiliary field.

 The parameters in \eqref{lam_res_WZ Ein} correspond to translations, local Lorentz transformations and the $U(1)$ gauge transformations.  From eq.~\eqref{eq: H E transf} we derive the full set of the transformation laws for the component fields:
\begin{equation}\label{Eins trans laws}
\begin{split}
   & \delta^* e_a^n = e^m_a \partial_m a^n - a^n \partial_n e^m_a
    + l_{[a}^{\;\;b]} e^n_b,
\\
    & \delta^* V^{n(ij)} = V^{m(ij)} \partial_m a^n - a^m \partial_m V^{n(ij)},
\\
    & \delta^* R_a = - a^n \partial_n R_a  - l_{[a}^{\;\;b]} R_b,
\\
     & \delta^* D_a = - a^n \partial_n D_a  - l_{[a}^{\;\;b]} D_b,
\\
    &\delta^* T_{(\alpha\beta)} 
    =
    -
     a^n \partial_n T_{(\alpha\beta)} 
     - l_{(\alpha}^{\;\;\gamma)}  T_{(\gamma\beta)}
    - l^{\;\;\gamma)}_{(\beta}  T_{(\alpha\gamma)} ,
\\
    &\delta^* B_a = -a^n \partial_n B_a - l_{[a}^{\;\;b]} B_b + e^m_a \partial_m b,
\\
    &\delta^* S^{(ij)} = -a^n \partial_n S^{(ij)} .
    \end{split}
\end{equation}
The obtained bosonic sector coincides with the classic results
\cite{Fradkin:1979cw, Fradkin:1979as, deWit:1979dzm}.

\subsection{Hypermultiplet coupling}

For the $\mathcal{N}=2$ Einstein supergravity multiplet in WZ gauge, we find the following component expansion of the differential operator in the hypermultiplet EoM:
	\begin{equation}
		\begin{split}
			\mathbb{D}^{++} + \dfrac{1}{2} \mathbb{\Gamma}^{++} =& \;
			\partial^{++} 
			-
            2i \theta^{+} \sigma^a\bar{\theta}^{+} 
			\check{\nabla}_a
			-
			\\&
            - (\theta^+)^2
            \left(
             \bar{T} - i \partial_5 \right)
            -
            (\bar{\theta}^+)^2
            \left(
              T + i \partial_5 \right)
            +
            \\&+
            \left\{+ (\theta^+)^2 \bar{\theta}^{+\dot{\alpha}} P^{\alpha}_{\dot{\alpha}} 
			+(\bar{\theta}^{+})^2\theta^{+\beta}
            T_{(\beta}^{\alpha)} 
            +
            (\bar{\theta}^+)^2 \theta^{+\alpha} T 
            \right\} \partial_{\alpha}^- 
            \\&+
			\left\{ - (\bar{\theta}^{+})^2\theta^{+\alpha} \bar{P}^{\dot{\alpha}}_{\alpha}  
			+
			(\theta^+)^2 \bar{\theta}^{+\dot{\beta}}             \bar{T}_{(\dot{\beta}}^{\dot{\alpha})}
            +
             (\theta^+)^2 \bar{\theta}^{+\dot{\alpha}}             \bar{T} 
            \right\} \bar{\partial}_{\dot{\alpha}}^-
            \\&+
			(\theta^+)^4 
            \left\{
            (S^{(ij)}
            + 
            V^{m(ij)} e_m^a B_a) \partial_5 
            +
            V^{m(ij)} \partial_m
            + 
            \dfrac{1}{2} \partial_m V^{m(ij)}
            \right\}
            u^-_i u^-_j .
		\end{split}
	\end{equation}
We have introduced a useful notation for the "check covariant derivative":
\begin{equation}\label{eq: check der}
	\check{\nabla}_a := 
	e_a^m \partial_m + \dfrac{1}{2} \partial_m e^m_a - \text{Im} P_a 
	+ B_a \partial_5.
\end{equation}		
	
As in the conformal case, we start with the most general component expansion for the hypermultiplet given by eq. \eqref{eq: hyp exp} with an additional exponential $x^5$-dependence of the hypermultiplet superfield:
\begin{equation}\label{eq: hyper x5}
    \mathbf{q}^+ (\zeta, u, x^5) = e^{im_qx^5} q^+(\zeta, u),
\end{equation}
where the real parameter $m_q$ corresponds to the hypermultiplet mass in the flat limit.

The hypermultiplet EoM is rewritten component-wise as a system of harmonic equations
\begin{equation}\label{EoM_system}
\begin{cases}
\partial^{++} F^{+}  = \; 0,
\\
\partial^{++} M^- - \left(  \bar{T} + m_q \right) F^{+} = 0, 
\\
\partial^{++}  N^- - \; \left( T - m_q \right) F^{+} = 0 ,
\\
\partial^{++}  A^-_a  -  \check{\nabla}_a F^+
			= 0,
\\            
\partial^{++}  K^{(-3)} + 2\left( \check{\nabla}_a + 2 \text{Im} P_a \right) A^{-a} \;
		+ (T+m_q) M^- + (\bar{T}-m_q) N^-
            \\
            \qquad +
            \left\{
            im_q \left(S^{(ij)}
    + 
    V^{m(ij)} e_m^a B_a\right) 
    +
    V^{m(ij)}\partial_m
            +
            \dfrac{1}{2} \partial_m V^{m(ij)}
            \right\}
            u^-_i u^-_j F^+ = 0
            .
\end{cases}
\end{equation}
This system is structurally identical to that in the conformal case, see eqs.
\eqref{eq: conf hyp system} and \eqref{eq: SU(2) contr}. 

\smallskip

The algebraic equations can be solved explicitly:
\begin{equation}
    \begin{cases}
    F^+(x, u) =  f^i(x) u^+_i,
    \\
    M^-(x,u) =  \left( \bar{T}(x) + m_q \right) f^i(x) u^-_i ,
    \\
    N^-(x, u)  = \; \left(T(x) - m_q \right) f^i(x) u^-_i ,
    \\
	A^-_a(x,u) =  \check{\nabla}_a f^i(x) u^-_i. 
    \end{cases}
\end{equation}
The last equation in system \eqref{EoM_system} splits into a non-trivial algebraic equation on $K^{(-3)}$:
\begin{equation}
    K^{(-3)}  = - \dfrac{1}{3}\left\{
        im_q  (S^{(ij)}
    + 
    V^{m(ij)} e_m^a B_a)
    +
    V^{m(ij)}\partial_m
            +
            \dfrac{1}{2} \partial_m V^{m(ij)}
            \right\} f^k u^-_{(i} u^-_{j} u^-_{k)},
\end{equation}
and a dynamical equation on the complex doublet of scalars:
		\begin{equation}
        \begin{split}
            &
       2 \left( \check{\nabla}_a + 2 \text{Im} P_a \right) \check{\nabla}^a f^i
       +
      2 (T\bar{T}+m^2)  f^i 
            +
            \\&
            -
            \dfrac{2}{3}
            \left\{
            im_q S^{(ij)}
    +
    V^{m(ij)}\left( \partial_m + im_q e^a_m B_a \right) 
            +
            \dfrac{1}{2} \partial_m V^{m(ij)}
            \right\}
            f_j 
        = 0.
        \end{split}
	\end{equation}
As a result, after eliminating the auxiliary fields, we obtain the expression for the hypermultiplet in the physical sector:
    \begin{equation}
    \begin{split}
			 q^+(\zeta_A, u)\Big|_{\text{phys}} =& \; f^i u^+_i
			+
            (\theta^+)^2 \left( \bar{T} + m_q \right) f^i u^-_i
            +
            (\bar{\theta}^+)^2 \left( T - m_q \right) f^i u^-_i
            +
            \\&+
			2 i \theta^{+} \sigma^a\bar{\theta}^{+} \check{\nabla}_a f^i u^-_i
            +
            (\theta^+)^4
            K^{(-3)} .
    \end{split}
	\end{equation}

Let us plug the central-charged hypermultiplet \eqref{eq: hyper x5} into the hypermultiplet superfield action~\eqref{eq: hyp action}. As in the conformal case, the algebraic equations of motion from \eqref{EoM_system} zero out most of the contributions:
    \begin{equation}\label{S eins start}
        \begin{split}
            -\mathbf{\tilde{q}}^+ \left( \mathbb{D}^{++} + \dfrac{1}{2} \mathbb{\Gamma}^{++} \right) \mathbf{q}^+ 
            \qquad
            \to
            \qquad
           - \mathbf{\tilde{q}}^+\Bigg|_{(\theta^+)^0} \left[ \left( \mathbb{D}^{++} + \dfrac{1}{2} \mathbb{\Gamma}^{++} \right)  \mathbf{q}^+ \right] \Bigg|_{(\theta^+)^4}, 
    \end{split}
    \end{equation}
    and the component hypermultiplet action in the physical sector takes the following form:
   \begin{equation}
   	\begin{split}
   	S_{hyp}\Big|_{\text{phys}} 
   	=
   	\int d^4x \Big[ &\check{\nabla}^a f^i \check{\nabla}_a \bar{f}_i
   	-
   	(T\bar{T} +m_q^2) f^i \bar{f}_i
   		\\&+
   	\frac{im_q}{3} S^{(ij)} f_i \bar{f}_j
   -
   	\frac{1}{6} V^{m(ij)} \left( D^{u(1)}_m \bar{f}_i f_j
   	-
   	\bar{f}_i D^{u(1)}_m f_j
   	 \right)
   	 \Big],
   	    \end{split}
   	  \end{equation}	 
   	  where we denote the U(1) covariant derivative as
   	  \begin{equation}
   	  	\begin{split}
   	  	&D^{u(1)}_m f_i := \left(  \partial_m 
   	  	+
   	  	im_q e_m^a B_a \right)  f_i,
   	  	\\
   	  	 &D^{u(1)}_m\bar{f}_i := \left(  \partial_m 
   	  	 -
   	  	 im_q e_m^a B_a \right)  \bar{f}_i.
   	  \end{split}	
   	   	  \end{equation}
 To see why these derivatives are covariant, let us consider the hypermultiplet infinitesimal transformation law \eqref{eq: gauge transformations} in the presence of the $x^5$-dependence \eqref{eq: hyper x5}. It leads to  
\begin{equation}
	\delta^* f^i = - a^m \partial_m f^i - \dfrac{1}{2}  ( \partial_m a^m) f^i - i m_q b f^i.
\end{equation}    
The last contribution here corresponds to the infinitesimal $U(1)$ transformations of the scalar doublet.
Now let us make a scalar field redefinition following the corresponding analysis in the conformal case:
\begin{equation}
    f^i = \sqrt{e} \phi^i \qquad
    \to
    \qquad  \delta^* \phi^i = - a^m \partial_m \phi^i - im_q  b\,\phi^i.
\end{equation}
Using the property
\begin{equation}
\check{\nabla}_a (\sqrt{e} \phi^i) 
=
\sqrt{e} e^m_a  \mathcal{D}_m \phi^i ,
\end{equation}
where we have introduced yet another covariant derivative
\begin{equation}
  \mathcal{D}_m :=  \partial_m - D_m
	+
	im_q B_m,
\end{equation}	
we finally obtain the component hypermultiplet action in the Einstein supergravity background
  \begin{equation}\label{Einst final action}
\boxed{	\begin{split}
		S_{hyp}\Big|_{\text{phys}} 
		=
		\int d^4x e\, \Big[ &g^{mn} \mathcal{D}_m \phi^i \mathcal{D}_n  \bar{\phi}_i
		-
		(T\bar{T} +m_q^2) \phi^i \bar{\phi}_i
		\\&+
		\frac{im_q}{3} S^{(ij)} \phi_i \bar{\phi}_j
		-
		\frac{1}{6} V^{m(ij)} \left( \mathcal{D}_m \bar{\phi}_i \phi_j
		-
		\bar{\phi}_i \mathcal{D}_m \phi_j
		\right)
		\Big].
	\end{split} }
\end{equation}
In contrast to the conformal case \eqref{eq: hyp conformal}, this action does not contain a non-minimal coupling to gravity and is not invariant under local $SU(2)$ gauge transformations.

\section{Discussion}\label{sec: dis}

In this paper, we have presented the component analysis of
$\mathcal{N}=2$ supergravities in the harmonic superspace formalism.
We have obtained the Wess–Zumino-type gauge with the proper component field redefinitions for the bosonic sectors of the prepotentials of both the conformal and Einstein supergravities. Also, we have used the superfield actions of the hypermultiplet coupled to each of the gravity types to demonstrate the efficiency of our approach for the component analysis, as well as to obtain some results of independent value. 
For ease of future reference and for the reader's convenience, here we summarize our final results for the conformal case and provide reference links to the analogous formulas obtained for the Einstein one. 

\smallskip

The analytic prepotentials of the conformal $\mathcal{N}=2$ supergravity in the WZ-type gauge in presence of a vector superfield $\mathcal{H}_{WZ}^{++5}$, which acts as a second conformal compensator and corresponds to the central charge-related coordinate $x^5$,
are summarized as follows:
\begin{subequations}
\begin{equation}
	\begin{split}
			&\mathcal{H}^{++m}_{WZ} =  -2i \theta^{+} \sigma^a {\bar{\theta}}^{+} e^m_a
			+ (\theta^+)^4 V^{m\,ij} u^-_i u^-_j, 
			\\
			&\mathcal{H}_{WZ}^{++\beta+} = +
			(\theta^+)^2 \bar{\theta}^{+\dot{\alpha}} (\sigma^a)^{\beta}_{\dot{\alpha}} \left(R_a 
			-
			\frac{1}{4}\epsilon^{abcd} \omega_{b|cd}  +
			\frac{i}{2} e_b^m \omega_m^{ab}
			\right)
			+
			(\bar{\theta}^+)^2 \theta^{+\alpha} T_{(\alpha}^{\;\beta)},
			\\
			&\mathcal{H}^{++\dot{\beta}+}_{WZ}  = -(\bar{\theta}^+)^2 \theta^{+\alpha} 
			(\sigma^a)^{\dot{\beta}}_{\alpha}
			\left(R_a -  \frac{1}{4}\epsilon^{abcd} \omega_{b|cd}  -
			\frac{i}{2} e_b^m \omega_m^{ab}
			\right)
			+
			(\theta^+)^2 \bar{\theta}^{+\dot{\alpha}} \bar{T}_{(\dot\alpha}^{\;\dot{\beta})},
			\\    
			&\mathcal{H}^{(+4)}_{WZ}  = (\theta^+)^4\left( D-\frac{1}{3} R - \frac{1}{36} V_n^{(ij)} V^n_{(ij)} \right),
            \\
            &\mathcal{H}_{WZ}^{++5} = \;i (\theta^+)^2 C(x) - i (\bar{\theta}^+)^2 \bar{C}(x)
		-
		2i \theta^{+}\sigma^a\bar{\theta}^{+} B_{a}(x)
       \\
       &\qquad\qquad\qquad\qquad\qquad\qquad\qquad +(\theta^{+})^4\big(S^{(ij)} (x) +
	    B^a(x)  V_a^{(ij)}(x)\big) u^-_i u^-_j.
	\end{split} 
\end{equation}
The residual gauge  superparameters take the form
\begin{equation}
		\begin{split}
			&\lambda^m_{WZ}  = a^n - 2 i \theta^{+} \sigma^a\bar{\theta}^{+} e^m_a \lambda^{(ij)} u^-_i u^-_j
			+
			(\theta^+)^4 \lambda^{(ij)} V^{m(kl)} u^-_{(i} u^-_j u^-_k u^-_{l)},,
			\\
			&\lambda^{+\beta}_{WZ}  = \frac{1}{2}\theta^{+\beta} [d+ir] + \theta^{+\alpha} l^{(\beta}_{\;\alpha)}
			+
			    \theta^{+\alpha} \lambda^{(ij)}  u^+_i u^-_j 
			- 
			i (\theta^+)^2 \bar{\theta}^{+\dot\beta} (\sigma^a)^\alpha_{\dot{\beta}}e^m_a \partial_m \lambda^{ij} u^-_i u^-_j
			\\& \qquad\qquad +
			(\theta^+)^2 \bar{\theta}^{+\dot\beta} P_{\dot{\beta}}^\alpha \lambda^{ij} u^-_i u^-_j 
			+
			(\bar{\theta}^+)^2 \theta^{+\beta} T_{(\beta}^{\alpha)} \lambda^{ij} u^-_i u^-_j
			, 
			\\
			&\bar{\lambda}^{+\dot{\beta}}_{WZ}  = 
			\frac{1}{2}\bar{\theta}^{+\dot{\beta}} [d- ir]
			+
			\bar{\theta}^{+\dot\alpha} \bar{l}^{(\dot{\beta}}_{\;\dot{\alpha})}+
			    \bar{\theta}^{+\dot\alpha} \lambda^{(ij)}  u^+_i u^-_j
			-
			i(\bar{\theta}^+)^2 {\theta}^{+\beta} (\sigma^a)^{\dot\alpha}_{{\beta}}e^m_a \partial_m \lambda^{ij} u^-_i u^-_j
			\\&   
			\qquad
			\qquad -
			(\bar{\theta}^+)^2 {\theta}^{+\beta} \bar{P}_{\beta}^{\dot\alpha} \lambda^{ij} u^-_i u^-_j
			+
			({\theta}^+)^2 \bar\theta^{+\dot\beta} \bar{T}_{(\dot\beta}^{\;\dot\alpha)} \lambda^{ij} u^-_i u^-_j ,     
			\\    
			&\lambda^{++}_{WZ}  = 
			2 i \theta^{+} \sigma^a \bar{\theta}^{+} e_a^m \partial_m d
			+
			\lambda^{(ij)} u^+_i u^+_j
			+
			2 i \theta^{+} \sigma^a \bar{\theta}^{+} e^m_a \partial_m \lambda^{(ij)} u^+_i u^-_j -
			\\&  \qquad\qquad  -
			(\theta^+)^4
			\bigg(
			\frac{1}{3} V^{n(ij)} \partial_n \lambda^{(kl)} 
			u^-_{(i} u^-_{j} u^-_{k} u^+_{l)}
			+
			\dfrac{1}{2} V^{n \; j)}_{\; (k} \partial_n \lambda^{(k l)} u^-_{l} u^-_{j}
			\bigg)
			-
			\\&\qquad\qquad 
			- (\theta^+)^4 \Big( \nabla^a \nabla_a - D + \frac{1}{36} V_n^{(ij)} V_{(ij)}^n  + \dfrac{R}{3} \Big) \lambda^{(ij)} u^-_i u^-_j,
            \\&
            \lambda^5_{WZ}
            =  b
	           +
               \mathcal{H}_{WZ}^{++5}  \lambda^{--}, 
	\end{split}
\end{equation}
where $P_a = R_a 
-
\frac{1}{4}\epsilon^{abcd} \omega_{b|cd}  +
\frac{i}{2} e_b^m \omega_m^{ab}$.
Note that the formulas are significantly simplified in the $SU(2)$-singlet sector, i.e. for $\lambda^{(ij)} = 0$.

These residual transformation parameters carry a clear physical interpretation: $a^n(x)$ parameterizes general coordinate transformations, $l_{ab}(x)$ accounts for local Lorentz rotations, $v^{(ij)}(x)$ generates local $SU(2)$ transformations, while $d(x)$ and $r(x)$ correspond to local dilatations and $U(1)_R$ chiral rotations respectively.
Their action on the $\mathcal{N}=2$ Weyl multiplet fields is given by the following transformation laws:
\begin{equation}
	\begin{split}
		& \delta^* e_a^n = e^m_a \partial_m a^n - a^n \partial_n e^m_a
		+ l_{[a}^{\;\;b]} e^n_b
		- d e_a^{n},
		\\
		& \delta^* R_a = - a^n \partial_n R_a - \frac{1}{2} e_a^n \partial_n r 
		+
		l_{[a}^{\;\;b]} R_b 
		-
		d  R_a,
		\\
		&\delta^* T_{(\alpha\beta)} 
		=
		-
		a^n \partial_n T_{(\alpha\beta)} 
		- \dfrac{1}{2} [d- i r]\,T_{(\alpha\beta)}
		- l^{\;\;\gamma)}_{(\beta}  T_{(\alpha\gamma)} - l_{(\alpha}^{\;\;\gamma)}  T_{(\gamma\beta)},
		\\
		&\delta^* D = -  a^n \partial_n D  -2dD,
		\\
		&  \delta^* V^{m(ij)} 
		=  
		-
		a^n \partial_n V^{m(ij)} 
		-2d V^{m(ij)} 
		-6  \partial^m \lambda^{(ij)}
		+
		\lambda^{i}_{k} V^{m (kj)} + \lambda^{j}_{k} V^{m (ik)}
		\\
        &\delta^* C = - a^n\partial_n C  - (d+ir) C,
			\\
			&\delta^* B_a =  e_a^m \partial_m b  - a^n\partial_n B_a
			+ l_{[a}^{\;\;b]} B_b
			- d B_a,
			\\
			&\delta^* S^{(ij)}  = - a^n\partial_n  S^{(ij)}  - 2 d S^{(ij)} 
			+
			2 \lambda^{(i}_kS^{j)k}.
	\end{split}
\end{equation}
\end{subequations}
The corresponding formulas for the Einstein case are given by eqs.
\eqref{eq: Eins WZ}, \eqref{lam_res_WZ Ein},  \eqref{Eins trans laws}.

The component expressions for the free hypermultiplet coupled to supergravity, and to a vector multiplet in the conformal supergravity case, prove to take the following form in the physical sector, for the conformal and Einstein supergravities respectfully:
\begin{subequations}
\begin{equation}
	\begin{split}
		q^+(\zeta_A, u)\Big|^{\text{conf}}_{\text{phys}} = \; & \sqrt{e} \phi^i(x) u^+_i + m_q(\theta^+)^2\sqrt{e}  C \phi^i u^-_i
		-
		m_q (\bar{\theta}^+)^2 \sqrt{e} \bar{C} \phi^i  u^-_i
        +
		2 i \theta^{+} \sigma^a \bar{\theta}^{+} \sqrt{e}  e_a^m \mathcal{D}_m^{(conf)}
		\phi^i(x) u^-_i -
        \\&
        - \dfrac{1}{3} (\theta^+)^4
		\left\{
		im_q  (S^{(ij)}
		+ 
		V^{m(ij)} B_m)
		+
		V^{m(ij)}\partial_m
		+
		\dfrac{1}{2} \partial_m V^{m(ij)}
		\right\} \sqrt{e} \phi^k  \; u^-_{(i} u^-_j u^-_{k)}
		,
	\end{split}
\end{equation}
    \begin{equation}
	\begin{split}
		q^+(\zeta_A, u)\Big|^{\text{Ein}}_{\text{phys}} =& \; \sqrt{e} \phi^i u^+_i
		+
		(\theta^+)^2\sqrt{e}  \left( \bar{T} + m_q \right) \phi^i u^-_i
		+
		(\bar{\theta}^+)^2 \sqrt{e} \left( T - m_q \right) \phi^i  u^-_i
	\\&	+
		2 i \theta^{+} \sigma^a\bar{\theta}^{+}
		\sqrt{e} e_a^m \mathcal{D}_m^{(Enst)}  \phi^i u^-_i
		\\&
	-\dfrac{1}{3}	(\theta^+)^4
	\left\{
	im_q  (S^{(ij)}
	+ 
	V^{m(ij)} e_m^a B_a)
	+
	V^{m(ij)}\partial_m
	+
	\dfrac{1}{2} \partial_m V^{m(ij)}
	\right\} \sqrt{e}\phi^k u^-_{(i} u^-_{j} u^-_{k)};
	\end{split}
\end{equation}
\end{subequations}
with the covariant derivatives defined as follows:
\begin{equation}
\begin{split}
&\mathcal{D}_m^{(conf)} \phi^i 
	=
	\left( \partial_m + im_q B_m  \right) \phi^i 
	+
	\frac{1}{6} (V_m)^i_{\;j} \phi^j,
    \\
    &\mathcal{D}_m^{(Einst)} \phi^i:=  (\partial_m - D_m
	+
	im_q B_m) \phi^i,
    \end{split}
    \end{equation}    
    where $D_m$ is a Lorenz vector field (not to be confused with a covariant derivative).
We have also obtained the corresponding component actions in physical sectors, see eqs.~\eqref{conf final action: vector} and~\eqref{Einst final action}.

\smallskip

 These results provide the basis for further analysis of the component reduction in a curved harmonic superspace. Below, we outline some related issues, which we intend to explore in the near future.

\smallskip

\noindent $\bullet$ \textit{Wess-Zumino-type gauge in fermionic  and mixed sector}

\smallskip

 We have demonstrated how to impose the Wess—Zumino-type gauge in the bosonic sector within the harmonic formalism. The analysis for the fermionic sector can be carried out analogously. In this case, at the zeroth order, one must keep track of the effect produced by the following parameters:
 \begin{equation}
 	\lambda^{+\alpha}_{(0)} = \epsilon^{\alpha i}(x) u^+_i,
 	\qquad
 		\bar{\lambda}^{+\dot{\alpha}}_{(0)} = \bar{\epsilon}^{\dot{\alpha} i}(x) u^+_i,
 		\qquad
 	\lambda^{++}_{(0)} =  \theta^{+\alpha} \eta_\alpha^i(x) u^+_i
 	+
 	\bar{\theta}^+_{\dot{\alpha}} \bar{\eta}^{\dot{\alpha}i}(x)u^+_i.
 \end{equation}	
We plan to present these results together with further applications of the component reduction in a separate work.

\smallskip

\noindent $\bullet$ \textit{Solution of zero-curvature equations }
\smallskip

The next step in the study of the component reduction is to solve the zero-curvature equations in WZ gauge, which  take the form
\begin{equation*}
	\begin{split}
	&\mathcal{N}=2\;\text{conformal supergravity:}
	\qquad
	[\mathfrak{D}^{++} - (\mathfrak{D}^{--} \mathcal{H}^{(+4)}) \mathcal{D}^0, \mathfrak{D}^{--} ] = \mathcal{D}^0,
\\
	&\mathcal{N}=2\;\text{Einstein supergravity:}
	\qquad \; \; \; 
	[\mathbb{D}^{++}, \mathbb{D}^{--}] = \mathcal{D}^0.
	\end{split}
\end{equation*}	
These equations define the connections in the negatively charged harmonic derivatives:
\begin{equation*}
	\begin{split}
	&\mathcal{N}=2\;\text{conformal supergravity:}
	\qquad
	\mathfrak{D}^{--} =  \partial^{--} + \mathcal{H}^{--m}\partial_m 
	+
	\mathcal{H}^{--\hat{\alpha}+} \partial^-_{\hat{\alpha}}
	+
		\mathcal{H}^{--\hat{\alpha}-} \partial^+_{\hat{\alpha}}
,
\\
	&\mathcal{N}=2\;\text{Einstein supergravity:}
	\qquad\;\;\;
	\mathbb{D}^{--} =  \partial^{--} + H^{--m}\partial_m 
	+
	H^{--\hat{\alpha}+} \partial^-_{\hat{\alpha}}
		+
	H^{--\hat{\alpha}-} \partial^+_{\hat{\alpha}}
	+
	H^{--5} \partial_5.
		\end{split}
\end{equation*}
The $\mathcal{H}^{--}$ superfields obtained this way are the building blocks for the full measure of harmonic superspace and the covariant derivatives. 
These results can then be applied to the study of the component reduction for a whole class of superfield theories, in particular, to the action of the $\mathcal{N}=2$ vector multiplet on the supergravity background \eqref{eq: vect mult}. 

\smallskip

\noindent $\bullet$ \textit{Normal coordinates in curved harmonic superspace}

\smallskip

The local expansion of the metric in Riemann normal coordinates
\begin{equation}
g_{mn}(x)=\eta_{mn}
-
\frac{1}{3}R_{mknl}(0)x^k x^l
-
\frac{1}{6}\nabla_p R_{mknl}(0)x^k x^l x^p+\cdots
\end{equation}
is a fundamental tool in the curved-space quantum field theory and mathematical physics, see e.g. \cite{Buchbinder:2021wzv}.
By eliminating the first derivatives of the metric at a given point (i.e., by setting
$\Gamma^\lambda_{\mu\nu}(0)=0$), this gauge reduces the covariant derivatives into the ordinary
partial derivatives locally, allowing for perturbative computations around the flat space.
The quadratic and cubic curvature terms in the series are indispensable for the
heat-kernel expansion, where they determine the Seeley—DeWitt coefficients and hence
the ultraviolet divergences of the effective action. Furthermore, the same expansion
underlies the derivation of the nonlocal form factors such as $R\,\Box^{-1}R$.

A similar normal gauge also exists in the Ogievetsky—Sokatchev geometric approach to supergravity \cite{Ogievetsky:1980de, Khudaverdian:1982nf}. By using this gauge, one can significantly simplify the construction of the torsion and curvature tensors in terms of the axial prepotential \cite{Ogievetsky:1980qm} and the derivation of the equations of motion in supergravity \cite{Ogievetsky:1980qp, Bandos:1985un}. It would be highly desirable for the benefit of future applications to develop a similar gauge in 
$\mathcal{N}=2$ harmonic supergravity.


\acknowledgments

It is a pleasure and an honour for N.Z. to thank E.~Ivanov for many fruitful discussions. 
\\
This work was partially supported by
the Foundation for the Advancement of Theoretical Physics and
Mathematics "Basis", grant \verb|#| 25-1-1-10-4 .

\appendix

\section{Notations and conventions}\label{app: notations}

Throughout the text, we use the notation of \cite{18}.
We  work in the analytic basis in the curved 
$\mathcal{N}=2$ superspace parametized by the coordinates
\begin{equation}
    \left\{x^m, \theta^{+\hat{\alpha}},  \theta^{-\hat{\alpha}}, u^{\pm}\right\},
    \qquad
\hat{\alpha}:= (\alpha, \dot{\alpha}). 
\end{equation}
The main advantage of the analytical basis is that the covariant spinor derivative $\mathcal{D}^+_{\hat{\alpha}}$ is shortened: $\mathcal{D}^+_{\hat{\alpha}} = \partial^+_{\hat{\alpha}}$.
Here we use the standard shorthand notation for the spinor derivatives:
\begin{equation}
\partial^+_{\hat{\alpha}}
: = \frac{\partial}{\partial \theta^{-\hat{\alpha}}},
\qquad
\partial^-_{\hat{\alpha}} : = \frac{\partial}{\partial \theta^{+\hat{\alpha}}}.
\end{equation}

The spinor and SU(2) isospinor indices are raised and lowered with the help of the
epsilon-symbols $\epsilon_{\alpha\beta}$, $\epsilon^{\alpha\beta}$, $\epsilon_{ij}$, $\epsilon^{ij}$, which satisfy the relations
\begin{equation}
	\epsilon_{\alpha\beta}
	=
	-\epsilon^{\alpha\beta},
	\qquad
		\epsilon_{\dot{\alpha}\dot{\beta}}
	=
	-\epsilon^{\dot{\alpha}\dot{\beta}}
	,
	\qquad
		\epsilon_{\alpha\beta} 	\epsilon^{\beta\gamma} = \delta_\alpha^\gamma,
		\qquad
			\epsilon_{\dot{\alpha}\dot{\beta}} 	\epsilon^{\dot{\beta}\dot{\gamma}} = \delta_{\dot{\alpha}}^{\dot{\gamma}}.
\end{equation}	

The partial harmonic derivatives are defined as
	\begin{equation}
		\partial^{++} := u^{+i} \frac{\partial}{\partial u^-_i},
		\qquad
		\partial^{--} := u^{-i} \frac{\partial}{\partial u^+_i},
		\qquad
		\partial^0 
		:= u^{+i} \frac{\partial}{\partial u^+_i} -   u^{-i} \frac{\partial}{\partial u^-_i}
	\end{equation}
	and satisfy the $su(2)$ algebra relations $[\partial^0, \partial^{\pm\pm}] = \pm 2 \partial^{\pm\pm}$, $[\partial^{++}, \partial^{--}] = \partial^0$. Moreover, they preserve the equality $u^{+i} u^-_i =1$.

In harmonic superspace, the \textit{tilde conjugation} is introduced as a combination of the ordinary complex conjugation and the antipodal map on the sphere. Its action is given by the rules
\begin{subequations}\label{eq: tilde conj}
\begin{equation}
		\widetilde{f^{(i_1\ldots i_n)}}
        =
        \overline{f^{(i_1\ldots i_n)}}
        = \bar{f}_{(i_1\dots i_n)}, 
        \quad
        \widetilde{\epsilon^{ij}}
        =
        \overline{\epsilon^{ij}}
        =
       - \epsilon_{ij};
\end{equation}        
\begin{equation}    
			 \widetilde{u^{\pm}_i} = u^{\pm i},\quad
		\widetilde{u^{\pm i} }
        = - u^{\pm}_i,
        \quad
        \doublewidetilde{u^{\pm}_i}  = -u^{\pm}_i;
\end{equation}        
\begin{equation}        
\widetilde{x^m} = x^m,
\quad			 \widetilde{\theta^\pm_\alpha} = \bar{\theta}^\pm_{\dot{\alpha}},
             \quad		\widetilde{\bar{\theta}^\pm_{\dot{\alpha}}} = - \theta^\pm_\alpha,
 \qquad
 \doublewidetilde{\theta^{\pm}_{\hat{\alpha}}} = - \theta^{\pm}_{\hat{\alpha}}.
\end{equation}
\end{subequations}
In particular, this leads to the conjugation properties
 $$\widetilde{f^{i}u^+_i} = \bar{f}_i u^{+i} = -\bar{f}^i u^+_i,
 \qquad
 \widetilde{f_i} = \widetilde{\epsilon_{ij} f^j}
 =
  - \epsilon^{ij} \bar{f}_i = - \bar{f}^i.
 $$
The analytic superspace is real with respect to this kind of conjugation.

\subsection{Spinor notations}

 In conventions of \cite{18},
\begin{equation}\label{eq: def sigma}
	 \sigma^a_{\alpha\dot{\alpha}} (\sigma^b)^{\dot{\alpha}\beta}  
	=
	\eta^{ab}\delta_\alpha^\beta - i(\sigma^{ab})_\alpha^\beta,
	\;\;\;\;\;\;\;\;\;\;\;\;\;\;
	(\sigma^a)^{\dot{\alpha}\alpha} \sigma^b_{\alpha\dot{\beta}} = \eta^{ab} \delta^{\dot{\alpha}}_{\dot{\beta}} - i(\bar{\sigma}^{ab})^{\dot{\alpha}}_{\dot{\beta}},
\end{equation}
so the sigma matrices with two antisymmetrized Lorentz indices are defined as
\begin{equation}\label{eq: ant sigma}
    \begin{split}
    &(\sigma^{[ab]})_{(\alpha \beta)} := \dfrac{i}{2} \left[ (\sigma^{a})_{\alpha \dot{\alpha}} (\sigma^{b})_\beta^{\dot{\alpha}} - (\sigma^b)_{\alpha \dot{\alpha}} (\sigma^{a})_\beta^{\dot{\alpha}} \right], 
    \\
    &(\bar{\sigma}^{[ab]})_{(\dot{\alpha} \dot{\beta})} := \dfrac{i}{2} \left[ (\sigma^{a})^{\alpha }_{\dot{\alpha}} (\sigma^{b})_{\alpha\dot{\beta}} - (\sigma^{b})^{\alpha }_{\dot{\alpha}} (\sigma^{a})_{\alpha\dot{\beta}} \right].
    \end{split}
\end{equation}
Under Hermitian conjugation these matrices satisfy the condition $	[(\sigma^{ab})_\alpha^\beta]^\dagger = (\bar{\sigma}^{ab})_{\dot{\alpha}}^{\dot{\beta}} $. They allow one to convert a pair of antisymmetric Lorentz indices into the spinor ones. Below we will show how exactly this works.

\medskip

The connection between the vector and the spinor indices is given by the relations
\begin{equation}\label{eq: Lorenz and spinor}	
A^{\alpha\dot{\alpha}} = \sigma^{\alpha\dot{\alpha}}_a A^a
	\qquad 
    \leftrightarrow
    \qquad
	A^a = \frac{1}{2}\sigma^a_{\alpha\dot{\alpha}} A^{\alpha\dot{\alpha}}.
\end{equation}
We define the Lorentz transformations for a field with Lorentz indices as
\begin{equation}
	A^a \to A^{\prime a} = \Lambda^a_{\;\;b} A^b,
	\qquad
		\Lambda^a_{\;\;b} = \delta^a_b 
	+
	l^{[a}_{\;\,\;b]},
\end{equation}
and for those with the spinor ones the definition is
 \begin{equation}
 	\begin{split}
 	\psi_\alpha \to 	\Lambda_\alpha^{\;\beta}  \psi_\beta,
 	\qquad
    &\Lambda_\alpha^{\;\beta} 
	=
	\delta_\alpha^\beta 
	+
	l_{(\alpha}^{\;\;\beta)},
\\
		\bar{\psi}_{\dot{\alpha}} \to 
 		\bar{\Lambda}_{\dot{\alpha}}^{\;\dot{\beta}}
 			\bar{\psi}_{\dot{\beta}}
	,
	\qquad
	&\bar{\Lambda}_{\dot{\alpha}}^{\;\dot{\beta}}
    =
	\delta_{\dot{\alpha}}^{\dot{\beta}} 
	+
	\bar{l}_{(\dot{\alpha}}^{\;\;\dot{\beta})}. 
	\end{split}
\end{equation}	
The relation between the vector and spinor indices  \eqref{eq: Lorenz and spinor} leads to the equality
\begin{equation}\label{eq: rel Lor and spin}
    \left( \sigma_b \right)_{\alpha\dot{\alpha}} \Lambda^b_{\;\;a} 
    = 
    \Lambda_{\alpha}^{\;\;\beta} \bar{\Lambda}_{\dot{\alpha}}^{\;\;\dot{\beta}} \left(\sigma_a \right)_{\beta\dot{\beta}}
    \quad
    \Rightarrow
    \quad
     l_{[a}^{\;\,b]} (\sigma_b)_{\beta\dot{\beta}}
=
    -l_{(\beta}^{\;\;\alpha)} (\sigma_a)_{\alpha\dot{\beta}}
    -\bar{l}_{(\dot{\beta}}^{\;\;\dot{\alpha})} (\sigma_a)_{\beta\dot{\alpha}},
\end{equation}
which allows one to relate the Lorentz transformation parameters with spinor and Lorenz indices to each other:
\begin{equation}\label{eq: Lor self dual}
	l_{(\alpha\beta)} = \frac{i}{4} l_{[ab]} \sigma^{ab}_{(\alpha\beta)},
	\qquad
	\bar{l}_{(\dot{\alpha}\dot{\beta})}
	=
	-
	\frac{i}{4}  
	l_{[ab]} \bar{\sigma}^{ab}_{(\dot{\alpha}\dot{\beta})},
	\qquad
	l_{[ab]}
	=
	-
	\frac{i}{2} l_{(\alpha\beta)} \sigma_{ab}^{(\alpha\beta)} 
	+
	\frac{i}{2} \bar{l}_{(\dot{\alpha}\dot{\beta})}
	\bar{\sigma}^{(\dot{\alpha}\dot{\beta})}_{ab}.
\end{equation}	

\subsection{Duality transformation}

For any 4D antisymmetric tensor $F_{ab} = -F_{ba}$ one can define the dual tensor ${}^\star F^{[ab]}$:
\begin{equation}\label{eq: duality transf}
	{}^\star F^{[ab]} := \frac{i}{2} \epsilon^{abcd} F_{[cd]}
    \quad
    \Leftrightarrow
	\quad
	F_{[ab]} = \frac{i}{2}\epsilon_{abcd} {}^\star F^{[cd]},
	\qquad
	{}^\star{}^\star F^{[ab]} = 	F^{[ab]}. 
\end{equation}	
Any such tensor can be decomposed into the self-dual and the anti-self-dual parts:
\begin{equation}
    F_{[ab]} = F_{ab}^+ + F_{[ab]}^-,
\end{equation}
which are defined as
\begin{equation}
	F^{\pm}_{ab} := \frac{1}{2} \left( F_{ab} \pm {}^\star F_{ab} \right) 
\end{equation}	
and satisfy the (anti-)self-duality conditions:
\begin{equation}
{}^\star F^\pm_{ab} = \frac{1}{2} \left( {}^\star F_{ab} \pm F_{ab} \right)
	=
	\pm F^{\pm}_{ab}.
\end{equation}

Under the duality transformation \eqref{eq: duality transf}, the antisymmetric sigma-matrices \eqref{eq: ant sigma} satisfy
\begin{equation}
	{}^\star \sigma^{ab} = \frac{i}{2} \epsilon^{abcd} \sigma_{cd} =   
	\sigma^{ab},
	\qquad
		{}^\star \bar{\sigma}^{ab} = \frac{i}{2} \epsilon^{abcd} \bar{\sigma}_{cd} =   
	- \bar{\sigma}^{ab}.
\end{equation}	
Consequently, these properties allow one to decompose any antisymmetric tensor into the $(1,0)$ and the $(0,1)$ spin-tensors:
\begin{equation}\label{eq: antisym exp}
	F_{[ab]} = F^+_{[ab]} + F^-_{[ab]}
	=
	-\frac{i}{2} \sigma_{[ab]}^{(\alpha\beta)} \mathcal{F}_{(\alpha\beta)}
	+
	\frac{i}{2} \bar{\sigma}_{[ab]}^{(\dot{\alpha}\dot{\beta})} \bar{\mathcal{F}}_{(\dot{\alpha}\dot{\beta})}.	
\end{equation}	
Vice versa, one can also express the spin-tensors as
\begin{equation}\label{eq: D and SD}
    \mathcal{F}_{(\alpha\beta)} = \frac{i}{4} \sigma^{ab}_{(\alpha\beta)} F^+_{ab}
    =
     \frac{i}{4} \sigma^{ab}_{(\alpha\beta)} F_{ab},
    \qquad \bar{\mathcal{F}}_{(\dot{\alpha}\dot{\beta})} 
    =
    - \frac{i}{4} \bar{\sigma}^{ab}_{(\dot{\alpha}\dot{\beta})} F^-_{ab}
    =
    - \frac{i}{4} \bar{\sigma}^{ab}_{(\dot{\alpha}\dot{\beta})} F_{ab}.
\end{equation}
Thus, decomposition \eqref{eq: Lor self dual} corresponds to the splitting into the self-dual and the anti-self-dual parts.

\subsection{Spin connection}\label{eq: spin con}

In our notations, the action of the infinitesimal local Lorentz transformations parameterized by the antisymmetric tensor $l_{[ab]}(x)$ is given by
\begin{equation}
    \delta_{Lor} A_a =  l_a^{\;\;b}(x)  A_b,
    \qquad
    \delta_{Lor} A^a = - l_b^{\;\;a}(x)  A^b
    =
     l_{\;\;b}^{a}(x)  A^b.
\end{equation}
In the tetrad formalism, the covariant derivative of a field carrying a Lorentz index is defined as
\begin{equation}
 \nabla_a A_b := e_a^n \partial_n A_b
    +
    e_a^n(\omega_n)^c_{\;\;b} A_c,
    \qquad
    \nabla_a A^b := e_a^n \partial_n A^b
    -
    e_a^n(\omega_n)^b_{\;\;c} A^c,
    \label{spin_cov_der}
\end{equation}
where we have introduced the spin connection $(\omega_n)^b_{\;\;c}$, which transforms according to the law
\begin{equation}\label{eq: omega Lor}
    \delta_{Lor}\, \omega_{m|ab}(x)
= 
 \partial_{m} l_{ab}(x)
+  l_{a}{}^{c}(x)\,\omega_{m|cb}(x)
+  l_{b}{}^{c}(x)\,\omega_{m|ac}(x).
\end{equation}
This ensures a homogeneous transformation law for the covariant derivatives.

The commutator of the covariant derivatives $\nabla_a = e^n_a \partial_n + \frac{1}{2} \omega_a^{\;\;bc} M_{[bc]}$ has the form
\begin{equation}\label{eq: com}
    [\nabla_a, \nabla_b] =   T_{ab}^c \nabla_c + \frac{1}{2} R_{ab}^{\;\;\;\;cd} M_{[cd]}.
\end{equation}
Imposing the torsion-free constraint
\begin{equation}
T_{ab}^c = C_{[ab]}^{\;\;\,c}  -  e_a^n \left( \omega_n\right)_b^{\;\;c} +  e_b^n \left( \omega_n\right)_a^{\;\;c}   = 0
\end{equation}
allows us to uniquely express the spin connection in terms of the tetrad field only. By cycling the indices, we obtain
\begin{equation}\label{eq: Lor Con}
\omega_{m|bc} = e_m^a \omega_{a|bc},
\qquad
    \omega_{a| bc} = \frac{1}{2} \left(C_{ab|c} +  C_{ac|b} -  C_{bc|a} \right). 
\end{equation}
Here, $C_{ab}^{\;\;\,c}$ denotes the anholonomy coefficients defined via the commutator of the tetrad frame vectors:
\begin{equation}
 [e_a^m \partial_m, e_b^n \partial_n] = C_{ab}^{\;\;\,c} e_c^n \partial_n,
 \qquad
 C_{[ab]}^{\;\;\,c} = \left( e_a^m \partial_m e^n_b - e_b^m \partial_m e^n_a\right)e_n^c.
\end{equation}

Equation \eqref{eq: com} provides the standard definition for
 the Riemann curvature tensor:
\begin{equation}
R_{mn}{}^{ab}
= \partial_m\omega_n{}^{ab}-\partial_n\omega_m{}^{ab}
+ \omega_m{}^{a}_{\;\;c}\,\omega_n{}^{cb}
- \omega_n{}^{a}{}_{c}\,\omega_m{}^{cb}.
\end{equation}
Scalar curvature is constructed from $R_{mn}^{\;\;\;\;ab
}$  as follows:
\begin{equation}\label{eq:scalarR}
R \;=\; e_a^{m} e_b^{n}\, R_{mn}{}^{ab}
=
e_a^{m} e_b^{n}\Big(\partial_m\omega_n{}^{ab}-\partial_n\omega_m{}^{ab}
+ \omega_m{}^{a}{}_{c}\,\omega_n{}^{cb}
- \omega_n{}^{a}{}_{c}\,\omega_m{}^{cb}\Big).
\end{equation}

\section{Gauge $SU(2)$ transformations}\label{eq: gauge SU(2)}

The gauge transformation law for an $SU(2)$ non-Abelian field is given by
\begin{equation}\label{eq: gauge su(2) field}
\delta A_m^a = \frac{1}{g}\,\partial_m \alpha^a - \varepsilon^{abc}\,\alpha^{b} A_m^{c},
\qquad a=1,2,3.
\end{equation}
To be able to compare this with the transformation law for field $V^{(ij)}_m$ from the $\mathcal{N}=2$ Weyl multiplet, it is convenient to switch to the matrix representation:
\begin{equation}
    \alpha = \alpha^a T^a,\qquad A_m = A_m^a T^a.
\end{equation}
For the fundamental (doublet) representation of \(SU(2)\), let the generators be
\[
(T^a)_i^{\;\;j} = \frac{(\sigma^a)_i^{\;\;j}}{2},\qquad a=1,2,3,
\]
where \(\sigma^a\) are the Pauli matrices.
Using the commutation relations for the \(SU(2)\) generators \([T^a, T^b] = i \varepsilon^{abc} T^c\) and multiplying Eq. \eqref{eq: gauge su(2) field} by \(T^{a}\), we can gather the terms into a matrix commutator:
\begin{equation}
\delta A_m = \frac{1}{g}\,\partial_m \alpha + i\,[\alpha, A_m],
\quad
\Leftrightarrow
\quad
\delta (A_m)_{i}^{\;\;j} = \frac{1}{g} \partial_m  \alpha_{i}^{\;\;j}
+
i \left( 
 \alpha_{i}^{\;\;k} (A_m)_{k}^{\;\;j} 
 -
 (A_m)_{i}^{\;\;k} \alpha_{k}^{\;\;j} 
\right).
\end{equation}

To establish a direct link between this standard expression and the transformation law of $V_m^{(ij)}$ from the Weyl multiplet, we use the invariant antisymmetric tensor \(\varepsilon _{ij}\) (\(\varepsilon_{12} = - \varepsilon^{12} = 1\)) as a metric to raise and lower the \(SU(2)\) doublet indices. Specifically, for any doublet \(X_{i}\), we define \(X^i = \varepsilon^{ij} X_j\). Contracting the generators with the antisymmetric tensor yields symmetric matrices:
\begin{equation}
(T^a)^{ij}\equiv \varepsilon^{ik} (T^a)^{\;\;j}_{k},
\qquad
(T^a)^{ij}=(T^a)^{ji},
\qquad
\overline{(T^a)^{ij}}
=
- T^a_{(ij)}
.
\end{equation}
Consequently, the gauge transformation law \eqref{eq: gauge su(2) field} for the $SU(2)$ field is rewritten in the form
\begin{equation}\label{eq: SU(2) form}
\begin{split}
    \delta A_m^{(ij)} & =
    \frac{1}{g} \partial_m \alpha^{(ij)} 
     - i \left( 
   A_m^{(ik)} \alpha_k^{\;\;j}
       +
   A^{(jk)}_m \alpha_k^{\;\;i} \right).
   \end{split}
\end{equation}

In the $SU(2)$ sector of the conformal supergravity multiplet, field $V_m^{(ij)}$ and the corresponding gauge parameter satisfy the following reality conditions:
\begin{equation}
	\overline{V_m^{(ij)}}
	=
	V_{m(ij)},
	\qquad
	\overline{\lambda^{(ij)}} 
	=
	\lambda_{ij}.
\end{equation}	
After the redefinitions
\begin{equation}
	V_m^{(ij)} = - 6i \mathcal{V}_m^{(ij)},
	\qquad
	\lambda^{(ij)}
	=
	i v^{ij},
\end{equation}	
the transformation law
\eqref{eq: su(2) vector} takes the form \eqref{eq: SU(2) form}. 


\end{document}